\documentclass[reprint,showkeys,  
preprintnumbers,amsmath,amssymb]{revtex4}
\usepackage{graphicx}
\usepackage{dcolumn}
\usepackage{bm}
\usepackage{color}

\numberwithin{equation}{section}

\usepackage{hyperref}
\begin{document}
\title{One-parameter families of supersymmetric isospectral potentials from Riccati solutions in function composition form}
%
\author{Haret C. Rosu}
\email{hcr@ipicyt.edu.mx}
\affiliation{IPICYT, Instituto Potosino de Investigacion Cientifica y Tecnologica,\\
Camino a la presa San Jos\'e 2055, Col. Lomas 4a Secci\'on, 78216 San Luis Potos\'{\i}, S.L.P., Mexico}
\author{Stefan C. Mancas}
\email{mancass@erau.edu}
\affiliation{Department of Mathematics, Embry-Riddle Aeronautical University, Daytona Beach, FL 32114-3900, USA}
\author{Pisin Chen}
\email{pisinchen@phys.ntu.edu.tw}
\affiliation{Leung Center for Cosmology and Particle Astrophysics (LeCosPA) and Department of Physics, National Taiwan University, Taipei 10617, Taiwan}

\medskip

\date{11 February 2014}
\begin{abstract}
In the context of supersymmetric quantum mechanics, we define a potential through a particular Riccati solution of the composition form $(F\circ f)(x) =F(f(x))$ and obtain a generalized Mielnik construction of one-parameter isospectral potentials when we use the general Riccati solution.
Some examples for special cases of $F$ and $f$ are given to illustrate the method.
An interesting result is obtained in the case of a parametric double well potential generated by this method, for which it is shown that the parameter of the potential controls the heights of the localization probability in the two wells, and for certain values of the parameter the height of the localization probability can be higher in the smaller well.
\\

\medskip

{\small Highlights: \\

Function-composition generalization of parametric isospectral potentials is presented.

Mielnik one-parameter family of harmonic potentials is obtained as a particular case.

Graphical discussion of regular and singular regions in the parameter space is given.}
\end{abstract}

\medskip

\begin{center} {\tt Ann. Phys. 343, 87-102 (2014)} \end{center}
\begin{center} {\tt arXiv:1310.6642v3} \end{center}

\maketitle

\section{Introduction}

Isospectral potentials in nonrelativistic quantum mechanics have been of much interest as one of the standard way to extend the class of exactly solvable  Schr\"odinger problems. For many decades in the past, the main activity in this area was through the factorization method \cite{FM0,FM1,FM2}, and it was only during the 1980's when the connections with the Darboux transformations have been fully used by mathematical physicists in the novel realm of supersymmetric quantum mechanics (SUSY QM). The origin of the latter is usually traced back to a paper by Witten \cite{witt} in which he used the example of a spin one half particle on the line as a simple realization of the supersymmetric algebra of bosonic and fermionic operators obeying commutation and anticommutation relations. Witten noticed that the diagonal $2\times 2$ Hamiltonian matrix having the spin up and spin down Hamiltonians as components, and the $2\times 2$ matrices of the factorization operators of the two Hamiltonians, also known as supercharges, i.e.,
$$
H=\left(\begin{array}{cc}
H_1 & 0\\
0 & H_2\end{array}
\right)~, \qquad Q^-=\left(\begin{array}{cc}
0 & 0\\
A & 0\end{array}
\right)~,  \qquad Q^+=\left(\begin{array}{cc}
0 & A^+\\
0 & 0\end{array}
\right)~,
$$
satisfy a closed superalgebra
$$
[H,Q^-]=[H,Q^+]=0~, \quad \{Q^-,Q^+\}=H~, \quad \{Q^-,Q^-\}=\{Q^+,Q^+\}=0~.
$$
The fact that the supercharges commute with $H$ means that $H_1$ and $H_2$ are isospectral. In quantum field theory, the supercharge operators are responsible for changing  bosonic degrees of freedom into fermionic ones and vice versa, however in quantum mechanics they are known as intertwinning operators that factorize the supersymmetric partner Hamiltonians, $H_1=A^+ A$ and $H_2=AA^+$, respectively. If however one `forgets' about this matrix representation, it is easy to see that SUSY QM is a reformulation of the factorization method already introduced through simple examples by Dirac and Schr\"odinger at the beginning of quantum mechanics. In fact, for the spin one half case, Witten wrote the expressions of the supercharges as matrix first order operators involving an unknown function $W(x)$ and also the matrix Hamiltonian in terms of $W^2$ and $W'$ but his focus was on the dynamical supersymmetry breaking issue and not on eigenvalue problems in quantum mechanics. Later, the SUSY QM authors noted that Witten's superpotential $W$ is the solution of coupled Riccati equations involving the two isospectral potentials, which was a new basic result for the factorization method and turned it towards the goal of finding isospectral partner potentials to exactly solvable known potentials. Since SUSY QM is essentially an algebraic scheme, it is convenient to shift the ground state energy of $H_1$ such that to eliminate the contribution to the energy of the zero point fluctuations. In this way, the energy spectra of the two partner Hamiltonians are semipositive definite. For $n>0$, the Schr\"odinger equation for $H_1$
\begin{equation}\label{i1}
H_1\phi_n=A^+ A\phi_n=\epsilon_n\phi_n
\end{equation}
can be written as a Schr\"odinger equation for the SUSY partner Hamiltonian $H_2$, i.e.,
\begin{equation}\label{i2}
H_2(A\phi_n)=AA^+ A\phi_n=\epsilon_n(A\phi_n)
\end{equation}
by applying the $A$ operator to the left in (\ref{i1}). Thus if one uses the wavefunctions $\varphi_n=A\phi_n$
in the eigenvalue problem $H_2\varphi_n=\varepsilon \varphi_n$ then this is isospectral to the eigenvalue problem for $H_1$, i.e., $\varepsilon_n=\epsilon_n$. Vice versa, by applying the operator $A^+$ to the left in the eigenvalue problem for $H_2$, i.e.,
\begin{equation}\label{i3}
A^+ H_2\varphi_n=A^+ \varepsilon \varphi_n~,
\end{equation}
one immediately gets by rearranging
\begin{equation}\label{i4}
H_1(A^+\varphi_n)= \varepsilon (A^+\varphi_n)
\end{equation}
and therefore by comparing with (\ref{i1}) shows that choosing the eigenfunctions of $H_1$ as $\phi_n=A^+\varphi_n$ leads to the isospectrality
$\epsilon_n=\varepsilon$. Now, since the operator $A$ ($A^+$) is a first-order differential operators it also destroys (creates) a node in the eigenfunction on which it acts. Since the ground state of $H_1$ is nodeless, it means that no normalizable eigenfunction of $H_2$ is obtained by applying $A$ to $\phi_0$. This is the physical meaning of the equation $A\phi_0=0$, which can be also used as the mathematical definition of the ground state of $H_1$, considered as nondegenerate, a case known as unbroken supersymmetry. As a consequence, the eigenvalues and normalized eigenfunctions of a pair of SUSY QM Hamiltonians are related by
$$
\varepsilon_n=\epsilon_{n+1}~, \quad \epsilon_0=0~,
$$
$$
\varphi_n=\frac{A\phi_{n+1}}{\sqrt{\epsilon_{n+1}}}~,
$$
$$
\phi_{n+1}=\frac{A^+ \varphi_n}{\sqrt{\epsilon_{n}}}~.
$$
On the other hand, from the strictly physical application standpoint it is fair to say that no real applications emerged from SUSY QM and only a few consequences are claimed, such as the value of $g=2$ for the $g$ factor of the electron in a uniform magnetic field as described by a Pauli Hamiltonian, the $Z/n$ dependence of the Coulomb energies in the radial Coulomb problem \cite{hr86}, and in nuclear physics where supersymmetry was shown to establish certain links among the spectroscopic properties of different nuclei \cite{bijker}. This is similar to what happens with the supersymmetric quantum field models for which the experimental evidence of supersymmetric partner particles is lacking despite considerable search effort along the last four decades which explains why the main focus is on supersymmetry breaking. In SUSY QM, this disappointing situation can be also understood intuitively by simply examining the plots of the supersymmetric pair of isospectral potentials $V_{1,2}$ based on the particular Riccati solution as given in this paper. One can notice that their shapes are identical which clearly points to serious difficulties, if not the impossibility, of detecting some difference between them through the associated conservative forces. However, the situation changes when one takes into account the one-parameter isospectral potentials obtained by employing the general Riccati solution. Again, from the plots of the examples given in the following, one can see that this class of potentials have deformed shapes with respect to the non-parametric supersymmetric potentials and thus one may have hopes to get some experimental evidence for this shape difference. This was one of the main motivations to write this paper on the three-decade-old topic of supersymmetric one-parameter potentials \cite{M,F,N,B1,B2,TR} that we also present in a new perspective. Perhaps the most interesting result we obtain herein is the case of an asymmetric double-well deformation of a quartic potential for which we show that the parameter of the potential controls in an interesting way the heights of the probability densities in the two wells.

\medskip

The goal of this paper is to examine the simplest supersymmetric constructions taking as the starting point the corresponding Riccati equations but in a new formulation in which the Riccati solution is expressed in a function composition form, say $F(f(x))$. There are at least two advantages of this new formulation. One is that it is more general since once $F$ is chosen all the particular $f(x)$ subcases can be grouped under the same $F$-type case. Second, it is also related to a new integrability condition proposed by Mak and Harko \cite{MH} for more general reduced Riccati equations than those occurring in SUSY QM.
It is clear that any integrability condition of the Riccati equations opens the doors to SUSY QM type models but this was not noticed by Mak and Harko. Thus, in section \ref{R}, we present the functional composition choice of the particular Riccati solution in the SUSY QM context and also describe along the same lines the construction of the one-parameter class of isospectral potentials based on the general Riccati solution which leads easily to a wealth of interesting solvable cases. In section \ref{E}, we provide several illustrative cases. The first case, for which we completely describe two particular $f(x)$ subcases, is directly related to the integrability condition of Mak and Harko \cite{MH} on which we also briefly comment.
Next, a Fresnel-like case is displayed despite not being very common in the quantum mechanical context followed by a case leading to parametric quartic double well potentials for which we note that the localization probability of the particle in the wells is controlled by the parameter. The last case is that of the constant potential, which is still interesting despite its simplicity. Section \ref{D} contains a discussion of the cases and some possible applications. There is also an appendix about the normalization issue of the parametric zero modes.

\section{Riccati equations and factorizations in quantum mechanics}\label{R}

As mentioned in the Introduction, Riccati equations in the reduced form, i.e., without the linear term, are known to be important in SUSY QM ever since Witten discussed the spin one half quantum mechanical model for the dynamical symmetry breaking of supersymmetry in quantum field theory \cite{witt}.

Indeed, one can start the SUSY QM construction with an initial Riccati equation of the form
\begin{equation}\label{mh1}
-\Phi'+\Phi^2=V_1-\epsilon
\end{equation}
which comes from the non-operatorial part of the factorization $(-D+\Phi)(D+\Phi)\Psi=0$ of a given  Schr\"odinger eigenvalue problem
 \begin{equation}\label{s1}
-\Psi''+(V_1-\epsilon)\Psi=0.
\end{equation}
Then, one introduces a second partner equation
\begin{equation}\label{mh2}
\Phi'+\Phi^2=V_2-\epsilon
\end{equation}
obtained from the non-operatorial part of the factorization $(D+\Phi)(-D+\Phi)\tilde{\Psi}=0$ of the isospectral partner problem
\begin{equation}\label{s2}
-\tilde{\Psi}''+(V_2-\epsilon)\tilde{\Psi}=0
\end{equation} to the given  Schr\"odinger equation.

For $\epsilon=0$, one can easily see that the relationships between the zero-energy (or zero mode) Schr\"odinger solution $\Psi_0$ and the Riccati solution $\Phi$ are the following ones
\begin{equation}\label{conn}
\Psi_0=e^{-\int^x \Phi(x')dx'}~, \qquad \Phi=-\frac{\Psi'_0}{\Psi_0}=-\frac{d}{dx}\log\Psi_0~.
\end{equation}
The nodeless feature of $\Psi_0$ ensures that the Riccati solution is regular which leads to a regular partner potential $V_2$ and wavefunctions $\tilde{\Psi}$. This is not the case if in the mappings (\ref{conn}) one uses any of the excited states of the discrete spectrum or an eigenfunctions of the continuous spectrum of the initial Schr\"odinger equation.

\medskip

Suppose now that we revert the scheme and start by giving a particular solution of the isospectral problem in the form
\begin{equation}\label{mh3}
\Phi_p(x)=F(f(x))~.
\end{equation}
Then, the potential $V_2$ is given by
\begin{equation}\label{mhF}
F'f'+F^2=V_2-\epsilon
\end{equation}
and one can use it to find the one-parameter family of potentials generated by using the general solution of (\ref{mh2})
\begin{equation}\label{mhG}
\Phi_g(x)=F(f(x))+\frac{1}{u(x)}~.
\end{equation}
Substituting this Ansatz for $\Phi_g$ in (\ref{mh2}), one gets
\begin{equation}\label{mh4}
F'f'-\frac{u'}{u^2}+F^2+2\frac{F}{u}+\frac{1}{u^2}=V_2-\epsilon~,
\end{equation}
which leads to
\begin{equation}\label{mh5}
-u'+2F(f)u+1=0~.
\end{equation}
The solution of the latter equation is
\begin{equation}\label{mh6}
u(x)=\frac{\gamma+\int_0^x\mu(x') dx'}{\mu(x)}~,
\end{equation}
where $\mu$ is the integrating factor given by
\begin{equation}\label{mh7}
\mu(x)=e^{-2\int_0^xF(f(x'))dx'}\equiv \Psi_0^2(x) ~.
\end{equation}
Thus,
\begin{equation}\label{mh8}
\Phi_g(x)=F(f(x))+\frac{\mu(x)}{\gamma+\int_0^x\mu(x') dx'}~.
\end{equation}
By subtracting \eqref{mh2} from \eqref{mh1} we obtain
\begin{equation}
V_{1\gamma}=V_2-2\Phi_g',
\end{equation}
which yields
\begin{equation}\label{mh9}
V_{1\gamma}(x)=V_1(x)-2\frac{d^2}{dx^2}\ln\left |\gamma+\int_0^x\mu(x')dx'\right |~.
\end{equation}
Equation (\ref{mh9}) defines a one-parameter family of potentials in which $V_1$ is included when $\gamma \rightarrow \infty$, each member of the family having the same SUSY partner potential $V_2$.

Since the Schr\"odinger  equation corresponding to the potential $V_{1\gamma}(x)$ can be transformed into the Riccati equation for $\Phi_g$ by using the transformation $\Phi_g=-\frac{\Psi'_{0\gamma}}{\Psi_{0\gamma}}$, then by means of (\ref{mh8}), (\ref{mh7}), and the first relationship in (\ref{conn}) one can get the zero mode wavefunction for each member of this parametric family of potentials in the form
\begin{equation}\label{mh9b}
\Psi_{0\gamma}(x)=\frac{\sqrt{\mu(x)}}{\gamma+\int_0^x\mu(x') dx'}\equiv \frac{\Psi_0(x)}{\gamma+\int_0^x\Psi_0^2(x') dx'} ~.
\end{equation}
The parameter $\gamma$ defines a range of existence of the family of regular potentials $V_{1\gamma}$ and eigenfunctions $\Psi_{0\gamma}$ from an initial value $\gamma_0$ up to a critical value $\gamma_s$ for which both  $V_{1\gamma}$, and $\Psi_{0\gamma}$ are unbounded.  Using \eqref{mhG} and \eqref{mh6} then
\begin{equation}\label{mh9c}
\gamma_0=\frac{1}{\Phi_g(0)-F(f(0))~}.
\end{equation}
Let us denote $\gamma(x)=\int_0^ {x}\mu(x') dx'$. Then, on the half line $(0, \infty)$, the potentials and eigenfunctions are unbounded at
\begin{equation}\label{mh9d}
\gamma_{s+}=-\gamma(x)~. 
\end{equation}
On the other hand, for the negative half axis $(-\infty,0)$, the unboundedness is reached at
\begin{equation}
\gamma_{s-}=\int_{-x}^0\mu(-x') dx'=-\gamma(-x).
\end{equation}
Moreover, when the integrating factor is an even function, for the negative half-axis $(-\infty,0)$, the unboundedness is reached at
\begin{equation}\label{mh9e}
\gamma_{s-,e}=\int_0^ {x}\mu(x') dx'\equiv \gamma(x)~.
\end{equation}
Thus, to avoid singularities in the even case, one should always choose $|\gamma|>|\gamma_s|$. A well-known example of an even integrating factor is Mielnik's parametric harmonic oscillator \cite{M}. The complete discussion of how the regions of regular and singular cases are obtained in the $\gamma$-parameter space can be done graphically and is deferred to section \ref{D}. Besides, for normalized zero modes, one should not take $\gamma$ in the interval $[-1,0]$, see the Appendix.

\medskip

Equation (\ref{mh9}) is a generalization from the compositional point of view of the formula for one-parameter isospectral potentials in SUSY QM as introduced by Mielnik \cite{M} in the particular case of the quantum harmonic oscillator. For the interpretation of the parametric potentials (\ref{mh9}) as the result of a sequence of two Darboux transformations, the reader is directed to the book chapter by Rosu \cite{r98}.

\medskip

\section{Examples}\label{E}

\begin{figure}[x!] 
\begin{center}
\resizebox*{0.3\textheight}{!}{
{\includegraphics{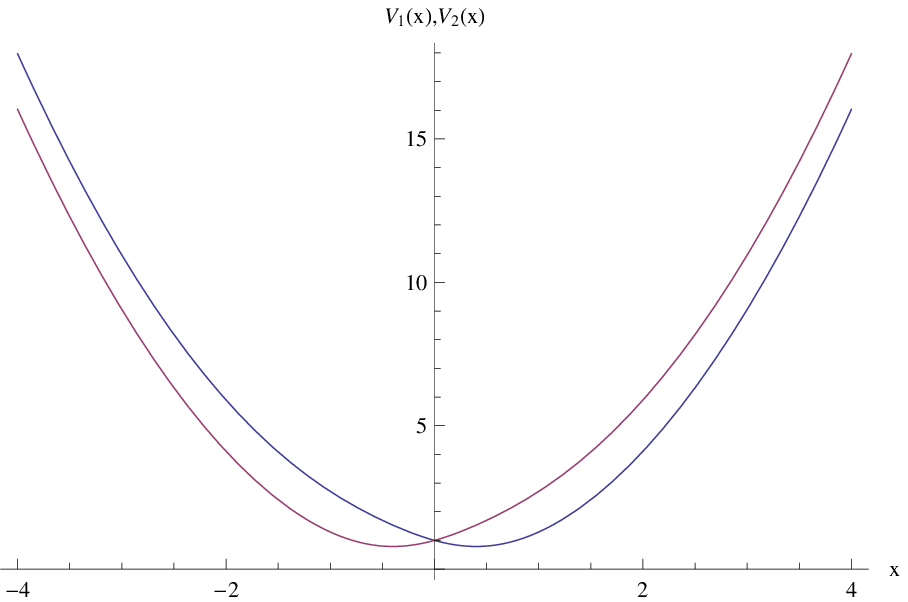}}}
\resizebox*{0.3\textheight}{!}{
{\includegraphics{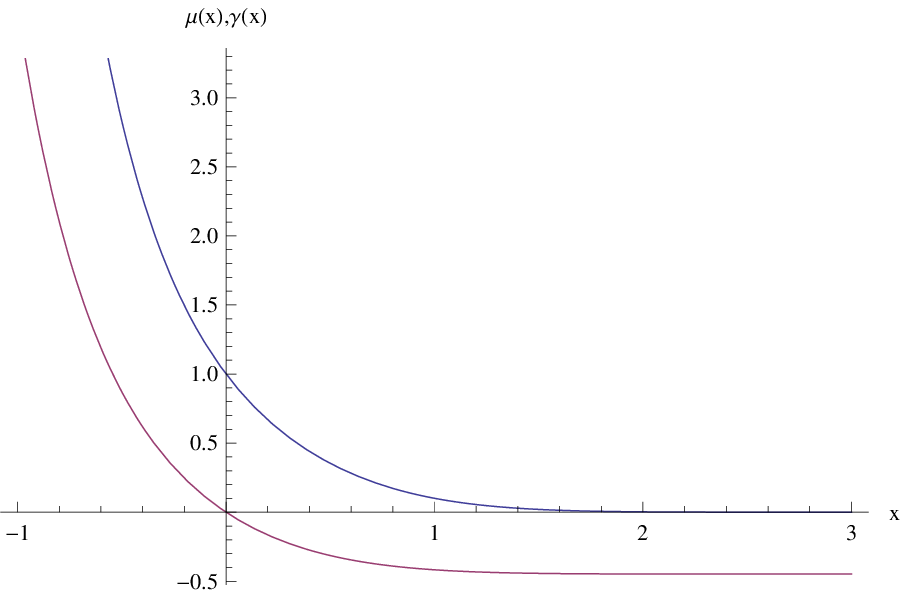}}}
\resizebox*{0.3\textheight}{!}{
{\includegraphics{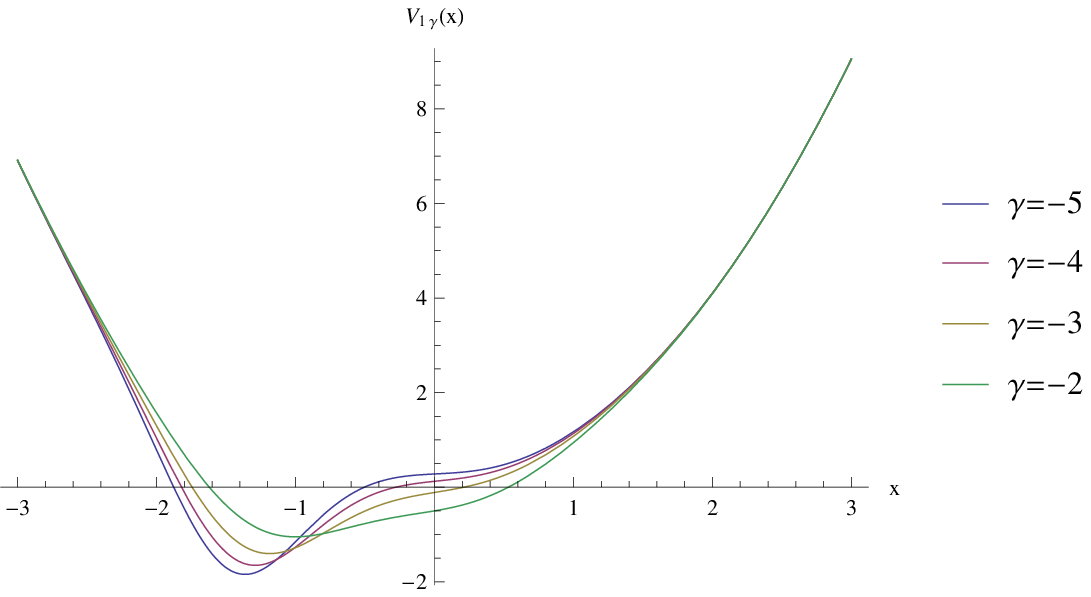}}}
\resizebox*{0.3\textheight}{!}{
{\includegraphics{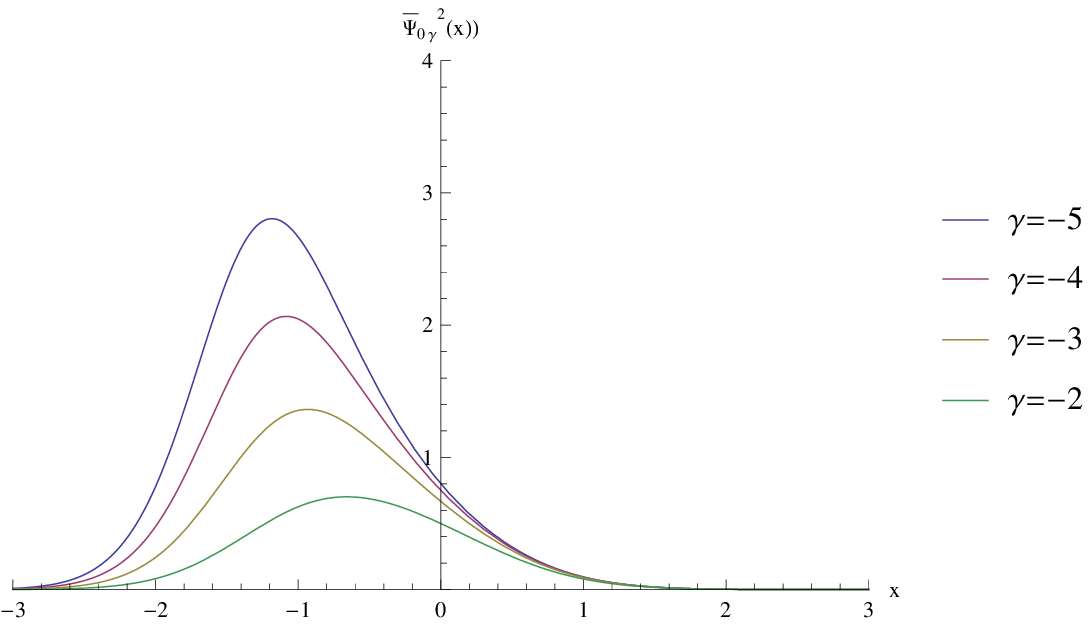}}}
\resizebox*{0.3\textheight}{!}{
{\includegraphics{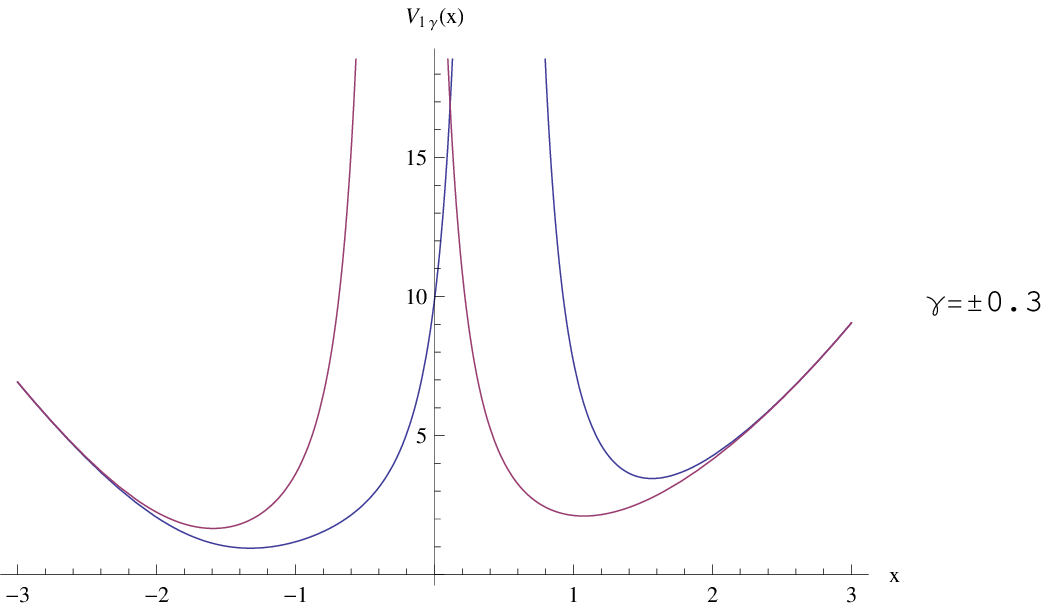}}}
\resizebox*{0.3\textheight}{!}{
{\includegraphics{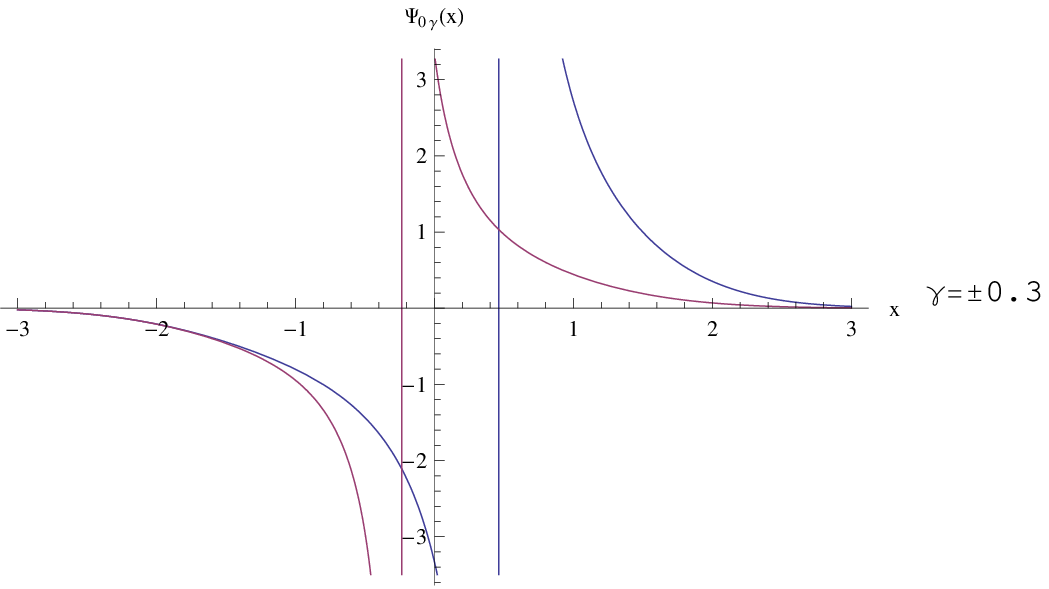}}}
\caption{\textsl{(color online). Case 1, subcase (a): asymmetric quadratic partner potentials, $V_1$ (blue) and $V_2$ (red); integrating factor $\frac{e^{-x\sqrt{x^2+1}}}{x+\sqrt{x^2+1}}$ (blue) and $\gamma(x;-1)$ (red);  one-parameter potentials, $V_{1\gamma}$, and normalized squared zero modes, $\overline{\Psi}_{0\gamma}^{2}$, for $\gamma=-5,-4,-3$ and $-2$; singular potentials and zero modes for $\gamma=\pm0.3$, red and blue, respectively.}}
\label{Set3}
\end{center}
\end{figure}

\begin{figure}[x!] 
\begin{center}
\resizebox*{0.3\textheight}{!}{
{\includegraphics{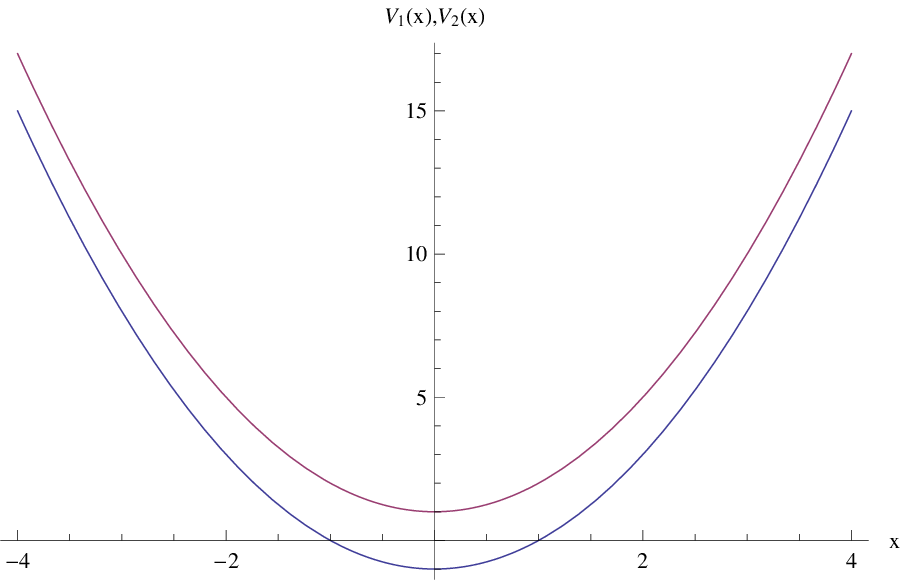}}}
\resizebox*{0.3\textheight}{!}{
{\includegraphics{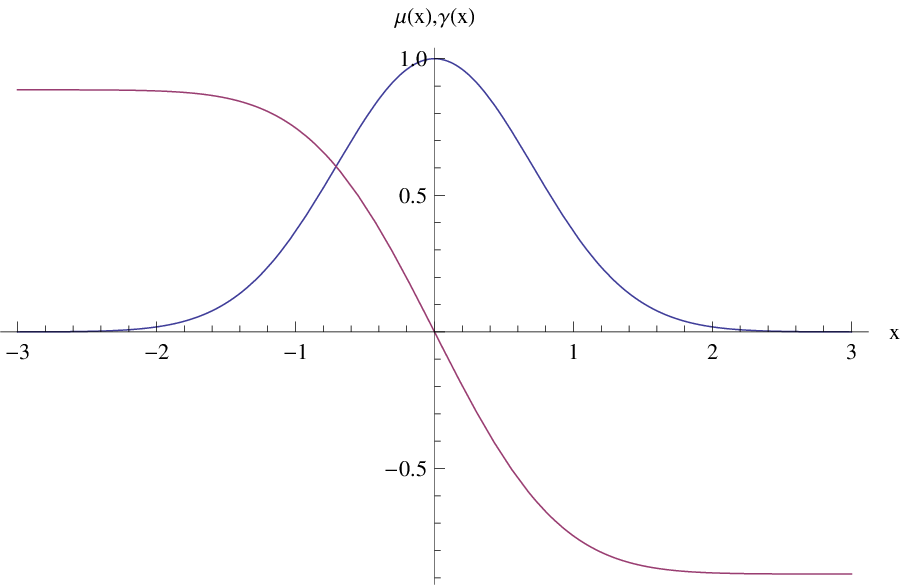}}}
\resizebox*{0.3\textheight}{!}{
{\includegraphics{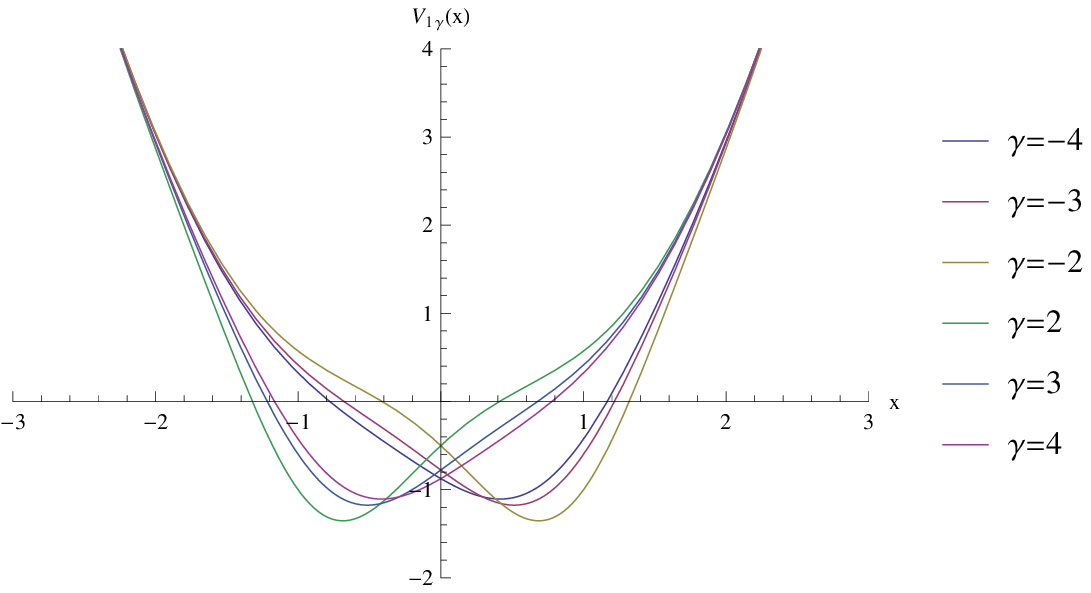}}}
\resizebox*{0.3\textheight}{!}{
{\includegraphics{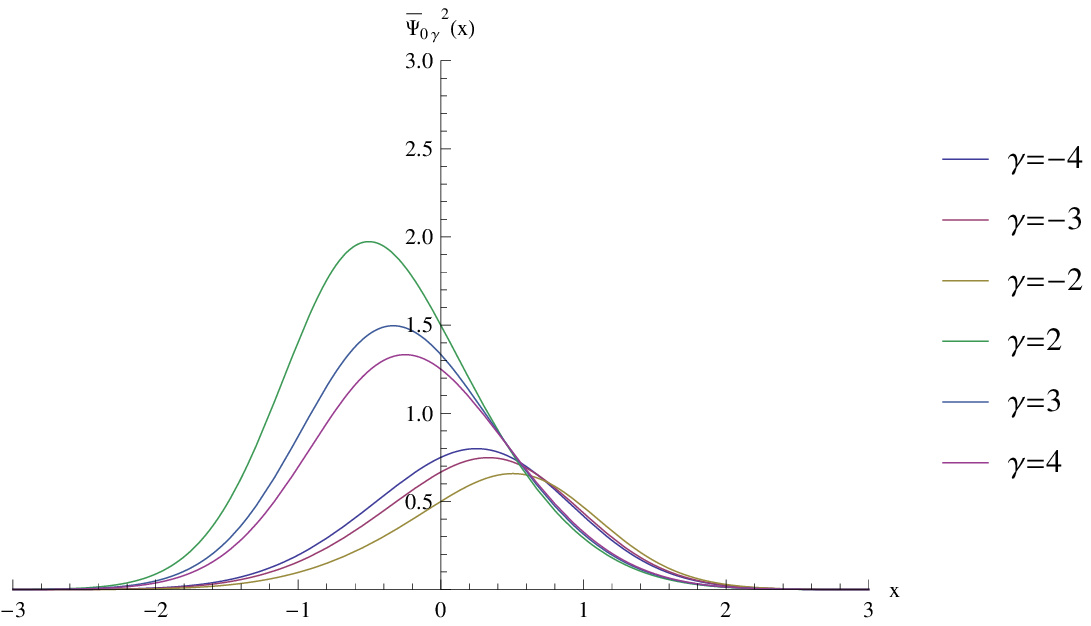}}}
\resizebox*{0.3\textheight}{!}{
{\includegraphics{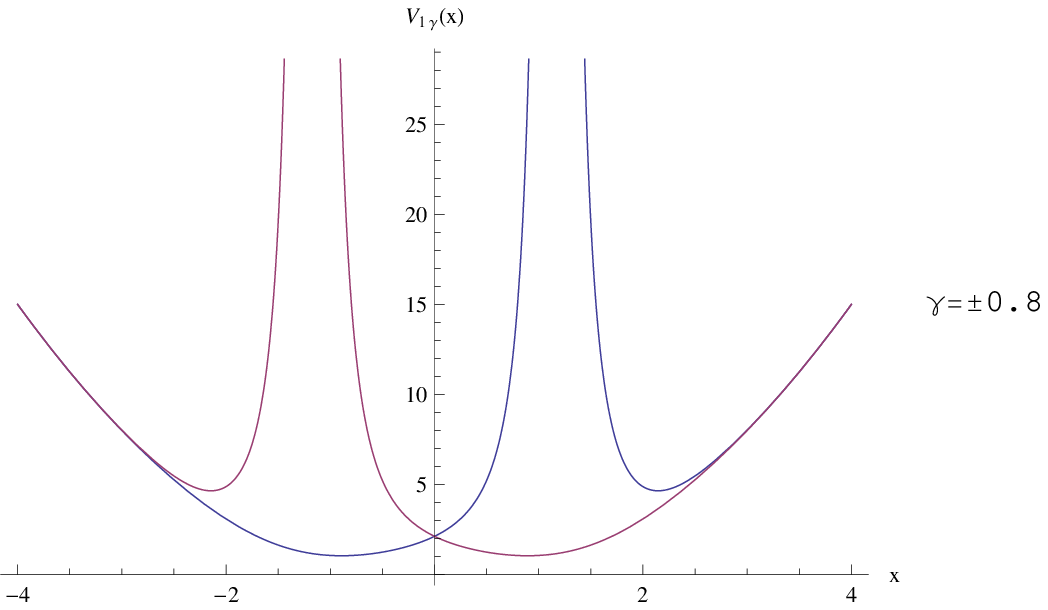}}}
\resizebox*{0.3\textheight}{!}{
{\includegraphics{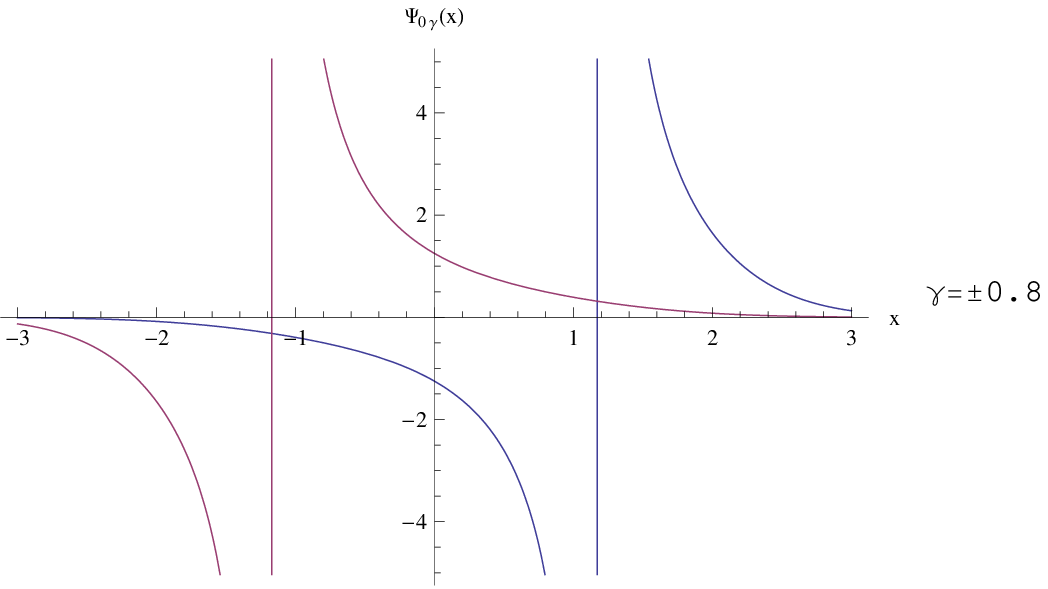}}}
\caption{\textsl{(color online). Case 1, subcase (b): symmetric quadratic, harmonic oscillator partner potentials, $V_1$ (blue) and $V_2$ (red);  integrating factor $e^{-x^2}$ (blue) and $\gamma(x)$ (red);  one-parameter potentials, $V_{1\gamma}$, and normalized squared zero modes, $\overline{\Psi}_{0\gamma}^{2}$, for $\gamma=-4,-3,-2,2,3$ and $4$; singular potentials and zero modes for $\gamma=\pm 0.8$, red and blue, respectively.}}
\label{Set4}
\end{center}
\end{figure}

\begin{figure}[x!] 
\begin{center}
\resizebox*{0.3\textheight}{!}{
{\includegraphics{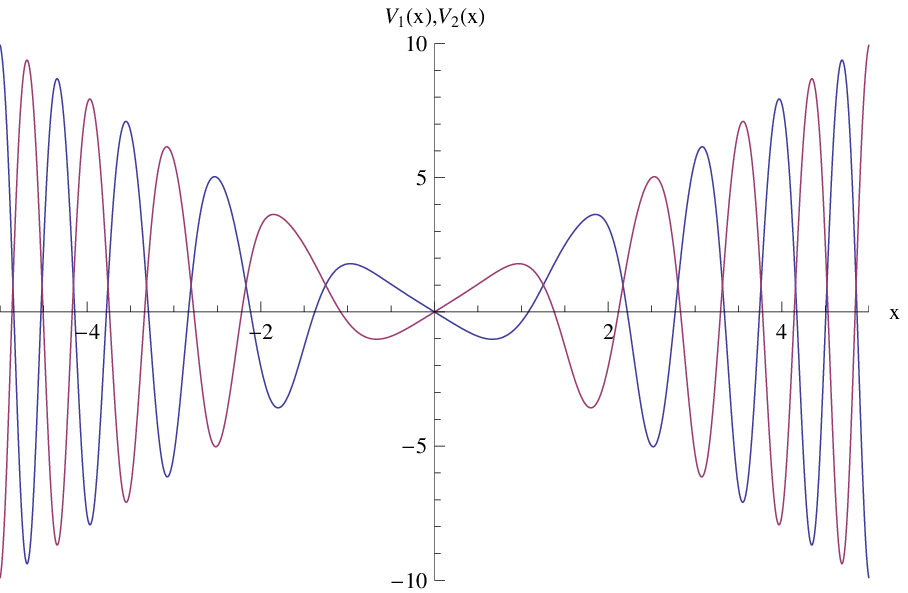}}}
\resizebox*{0.3\textheight}{!}{
{\includegraphics{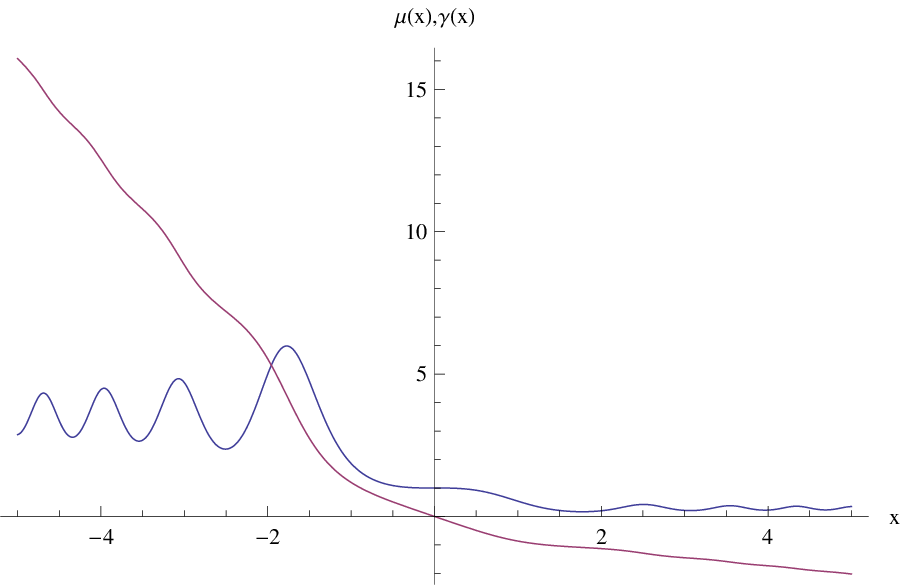}}}
\resizebox*{0.3\textheight}{!}{
{\includegraphics{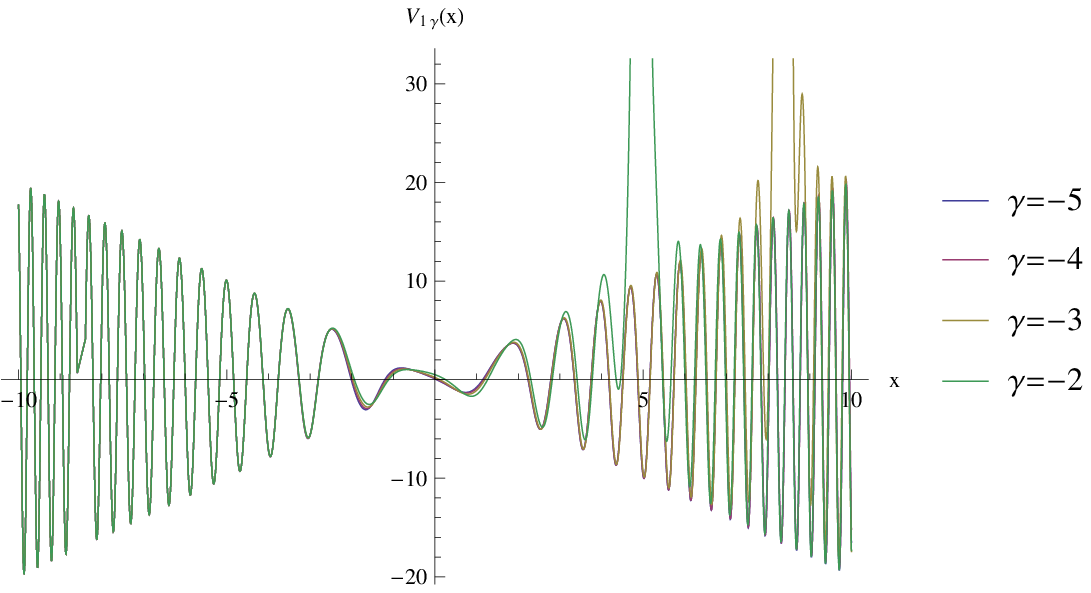}}} 
\resizebox*{0.3\textheight}{!}{
{\includegraphics{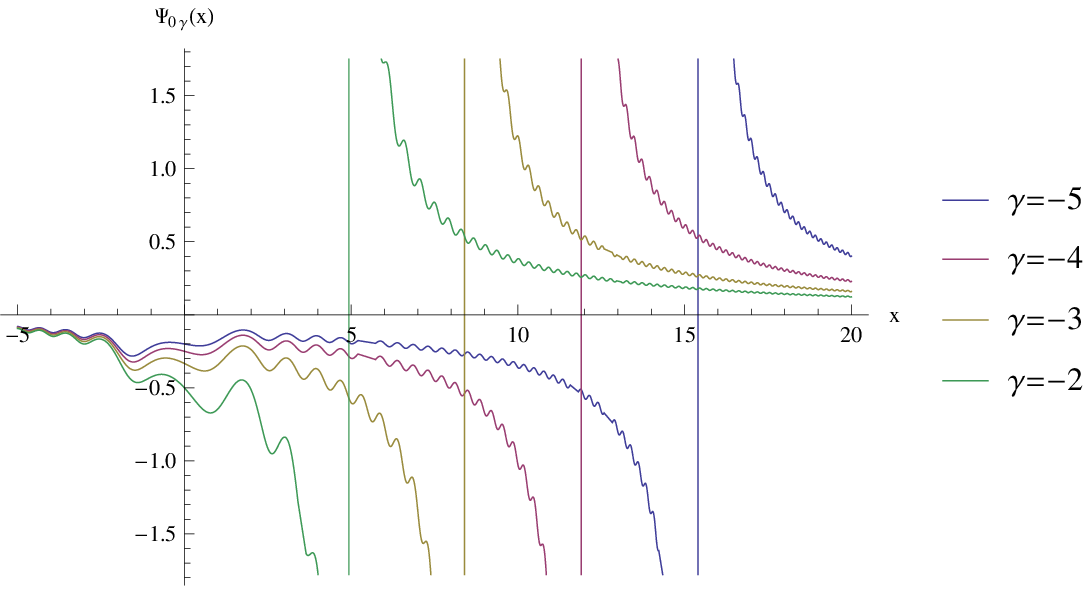}}}
\caption{\textsl{(color online). The Fresnel-like case: The partner potentials, $V_1$ (blue) and $V_2$ (red);  integrating factor $e^{-\sqrt{2\pi} S\left(\sqrt{\frac{2}{\pi}}x\right)}$ and its integral $\gamma(x)$, blue and red, respectively;  one-parameter potentials, $V_{1\gamma}$, and zero modes, $\Psi_{0\gamma}$, for $\gamma=-5,-4, -3$ and $-2$.}}
\label{Set5}
\end{center}
\end{figure}

\begin{enumerate}


%
%

\item \underline{$F(x)=\sqrt{x}$; $f(x)=a_nx^n+a_0$, $a_n \ne 0$}.

The square root type of particular Riccati solution has been recently proposed by Mak and Harko \cite{MH} in the disguised form of an integrability condition for Riccati equations of the form:
\begin{equation}\label{makh1}
y'=a(x)+c(x)y^2~.
\end{equation}
Their integrability condition states that if the coefficients $a(x)$ and $c(x)$ satisfy the condition
\begin{equation}\label{makh2}
\pm \frac{d}{dx}\left(\sqrt{f(x)/c(x)}\right)=a(x)+f(x)~,
\end{equation}
where $f(x)$ is an arbitrary function, then the general solution of the reduced Riccati equation (\ref{makh1}) can be obtained by quadratures.
However, for the standard quantum mechanics, $a(x)=-(V_{1,2}(x)-\epsilon)$ and $c(x)=1$ and then their integrability condition simply states that it is possible to get the general Riccati solution when the particular solution is given in the form $F=\sqrt{f}$ as described in the previous section.
For the power-law $f(x)$ (sub)case, we get
\begin{align}\label{power}
&V_{2,1}(x)=a_nx^n+a_0\pm \frac{na_nx^{n-1}}{2\sqrt{a_nx^{n}+a_0}}~,\\
&\mu(x)=e^{-\frac{2x\sqrt{a_nx^n+a_0}(2+n \,{}_2F_1(1,\frac{1}{2}+\frac{1}{n},1+\frac{1}{n};-\frac{a_nx^n}{a_0})}{2+n}}~,\\
&V_{1\gamma}(x)=a_nx^n+a_0-\frac{na_nx^{n-1}}{2\sqrt{a_nx^{n}+a_0}}
+\frac{4e^{-\frac{2x\sqrt{a_nx^n+a_0}(2+n{}_2F_1(1,\frac{1}{2}+\frac{1}{n},1+\frac{1}{n};-\frac{a_nx^n}{a_0}))}{2+n}}\sqrt{a_nx^n+a_0}}{\gamma+\int_0^x e^{-\frac{2x'\sqrt{a_nx'^n+a_0}(2+n{}_2F_1(1,\frac{1}{2}+\frac{1}{n},1+\frac{1}{n};-\frac{a_nx'^n}{a_0}))}{2+n}}dx'}+ \nonumber\\
&+\frac{2e^{-\frac{4x\sqrt{a_nx^n+a_0}(2+n{}_2F_1(1,\frac{1}{2}+\frac{1}{n},1+\frac{1}{n};-\frac{a_nx^n}{a_0}))}{2+n}}}{\left(\gamma+\int_0^x e^{-\frac{2x'\sqrt{a_nx'^n+a_0}(2+n{}_2F_1(1,\frac{1}{2}+\frac{1}{n},1+\frac{1}{n};-\frac{a_nx'^n}{a_0}))}{2+n}}dx'\right)^2}~,\\
&\Psi_{0\gamma}(x)=\frac{e^{-\frac{x\sqrt{a_nx^n+a_0}(2+n{}_2F_1(1,\frac{1}{2}+\frac{1}{n},1+\frac{1}{n};-\frac{a_nx^n}{a_0}))}
{2+n}}}{\gamma+\int_0^x
e^{-\frac{2x'\sqrt{a_nx'^n+a_0}(2+n\,{}_2F_1(1,\frac{1}{2}+\frac{1}{n},1+\frac{1}{n};-\frac{a_nx'^n}{a_0}))}{2+n}}dx'}~.
\end{align}

Unfortunately, because of the occurrence of the hypergeometric function ${}_2F_1$ in the exponent, the integrals can be performed analytically only in particular cases. We will discuss the reduction to the inverse hyperbolic function, and the error function.
\medskip

\begin{enumerate}

\item \underline{Subcase: $n=2$, $a_2=1$, $a_0=1$}.
In this particular subcase, one gets the following expressions:
\begin{align}\label{211}
&V_{2,1}(x)=x^2+1\pm \frac{x}{\sqrt{x^2+1}}~,\\
&\mu=\frac{e^{-x\sqrt{x^2+1}}}{x+\sqrt{x^2+1}}~,\\
&V_{1\gamma}(x)= x^2+1-\frac{x}{\sqrt{x^2+1}}+\frac{4\sqrt{x^2+1}e^{-x\sqrt{x^2+1}}}{(x+\sqrt{x^2+1})(\gamma+\int_0^x\frac{e^{-x'\sqrt{x'^2+1}}}
{x'+\sqrt{x'^2+1}}dx')}+
\frac{2e^{-2x\sqrt{x^2+1}}}{\left((x+\sqrt{x^2+1})(\gamma+\int_0^x\frac{e^{-x'\sqrt{x'^2+1}}}
{(x'+\sqrt{x'^2+1})}dx')\right)^2}~,\\
&\Psi_{0\gamma}(x)=\frac{e^{-\frac{1}{2}x\sqrt{x^2+1}}}{\sqrt{x+\sqrt{x^2+1}}\left(\gamma+\int_0^x\frac{e^{-x'\sqrt{x'^2+1}}}
{x'+\sqrt{x'^2+1}}dx'\right)}~.
\end{align}

Their plots are presented in Fig.~(\ref{Set3}). The critical $\gamma_s$ is given by
\begin{equation}\label{gs2}
\gamma_s=-\int_0^\infty \frac{e^{-x'\sqrt{x'^2+1}}}{x'+\sqrt{x'^2+1}}dx'\approx -0.44779~.
\end{equation}
For this case, we use $\gamma(x;-1)=\int_{-1}^{x}\mu(x')dx'$ for a sufficiently big $x$ as the equivalent of $\gamma_s$ and conclude graphically that regular potentials and zero modes occur for $\gamma<-0.44779$. If one requires normalized zero modes then one should take $\gamma<-1$.

\item \underline{Subcase: $n=2$, $a_2=1$, $a_0=0$}.
One can easily check that this is the well-known case of the quantum harmonic oscillator introduced by Mielnik \cite{M}.
Indeed, we have
\begin{align}\label{210}
&V_{2,1}(x)=x^2\pm 1~,\\
&\mu(x)=e^{-x^2}~,\\
&V_{1\gamma}(x)=x^2-1+\frac{4xe^{-x^2}}{\gamma +\frac{\sqrt{\pi}}{2}{\rm erf}(x)}+\frac{2e^{-2x^2}}{\left(\gamma +\frac{\sqrt{\pi}}{2}{\rm erf}(x)\right)^2}~,\\
&\Psi_{0\gamma}(x)=\frac{e^{-\frac{x^2}{2}}}{\gamma+\frac{\sqrt{\pi}}{2}{\rm erf}(x)}~, \label{210-4}
\end{align}
where erf is the error function, ${\rm erf}(x)=\frac{2}{\sqrt{\pi}}\int_0^x e^{-t^2}dt$.

Plots for this subcase are presented in Fig.~(\ref{Set4}). The critical parameter is that obtained by Mielnik \cite{M}
\begin{equation}\label{gs3}
|\gamma_s|=\int_0^ \infty e^{-x'^2}dx'=\frac{\sqrt \pi}{2}\approx 0.886227.
\end{equation}
The latter condition for regularity can be obtained graphically from the second plot in Fig.~(\ref{Set4}).
However, if normalized zero modes are required the valid range is $\gamma \in (-\infty, -1)\cup (0.886227,\infty)$.

\end{enumerate}

\item \underline{$F(x)=\sin x$; $f(x)=x^2$}.

For this periodic Riccati case the main quantities are the following:
\begin{align}\label{mhho-s0}
&V_{2,1}(x)=\sin^2(x^2)\pm 2x\cos(x^2)~,\\
&\mu(x)=e^{-\sqrt{2\pi} S\left(\sqrt{\frac{2}{\pi}}x\right)}~,\\
&V_{1\gamma}(x)=\sin^2(x^2)- 2x\cos(x^2)
+\frac{4e^{-\sqrt{2\pi}S\left(\sqrt{\frac{2}{\pi}}x\right)}\sin(x^2)}{\gamma +\int_0^x e^{-\sqrt{2\pi}S\left(\sqrt{\frac{2}{\pi}}x'\right)}dx'}+
\frac{2e^{-2\sqrt{2\pi}S\left(\sqrt{\frac{2}{\pi}}x\right)}}{\left(\gamma +\int_0^x e^{-\sqrt{2\pi}S\left(\sqrt{\frac{2}{\pi}}x'\right)}dx'\right)^2}~,\\
&\Psi_{0\gamma}(x)=\frac{e^{-\sqrt{\frac{\pi}{2}}S\left(\sqrt{\frac{2}{\pi}}x\right)}}{\gamma+\int_0^x
e^{-\sqrt{2\pi}S\left(\sqrt{\frac{2}{\pi}}x'\right)}dx'}~.
\end{align}

In these equations, the $S$ function is the Fresnel sine integral $S(x)=\int_{0}^{x}\sin \frac{\pi x'^2}{2}dx'$.
The plots corresponding to this periodic case are displayed in Fig.~(\ref{Set5}).
Due to the oscillatory properties of the Fresnel integral, for $x\leq 0 \rightarrow S(x)\leq 0$
\begin{equation}
\int_0^{\infty}e^{-\sqrt{2\pi} S\left(\sqrt{\frac{2}{\pi}}x'\right)} dx'=\int_0^{\infty}e^{-2 S(x')}dx'
\end{equation}
 is divergent since $e^{-2S(x)} \geq 1$. On the other hand, when $x>0$, $S(x)$ has an upper bound $M$, such that for $0 \leq S(x) \leq M<1$, then  $0<L=e^{-2M} \leq e^{-2S(x)} <1$, so $\int_0^{\infty}e^{-2 S(x')}dx'$ is still divergent.
 Here, $L \approx 0.167016$.



\item \underline{$F(x)=x^2-1$; $f(x)=x-1$}.

This is a case in which the supersymmetric partner potentials are quartic anharmonic potentials and the one-parameter family of isospectral potentials are asymmetric double well potentials. The basic quantities for this case are the following:
\begin{align}\label{E1}
&V_{2,1}(x)=x^4-4x^3+4x^2\pm 2x\mp 2=[x(x-2)]^2\pm 2x\mp 2~,\\
&\mu(x)=e^{-\frac{2}{3}x^2(x-3)}~,\\
&V_{1\gamma}(x)=[x(x-2)]^2-2x+2+\frac{4e^{-\frac{2}{3}x^2(x-3)}(x^2-2x)}{\gamma+\int_0^xe^{-\frac{2}{3}x'^2(x'-3)}dx'}+
\frac{2e^{-\frac{4}{3}x^2(x-3)}}{\left(\gamma+\int_0^xe^{-\frac{2}{3}x'^2(x'-3)}dx'\right)^2}~,\\
&\Psi_{0\gamma}(x)=\frac{e^{-\frac{x^2}{3}(x-3)}}{\gamma+\int_0^xe^{-\frac{2}{3}x'^2(x'-3)}dx'}~.
\end{align}
Plots of all these functions confirming the theoretical construction are presented in Fig.~(\ref{SetQ}).
The value of $\gamma_s$ is given by
\begin{equation}\label{gs-q}
\gamma_s=-\frac \pi 3 2^{\frac 2 3} e^{\frac 4 3}\mathrm{Bi}(2^{\frac2 3})-{}_2F_2\Big(\frac 1 2,1; \frac 2 3, \frac 4 3; \frac 8 3\Big) \approx -19.3694~, 
\end{equation}
where $\mathrm{Bi}(x)$ is the Airy function of the second kind and ${}_2F_2$ is a hypergeometric series with the first two Pochhammer symbols in the numerator and the last two in the denominator.

\item \underline{$F(x)=c$}.

This is the simplest possible case and corresponds to the constant potential. We have included it here because we want to draw attention on some interesting similarities with the case 1(a), see below.
The main quantities are the following:
\begin{align}\label{mhho-s0}
&V_{2,1}(x)=c^2~,\\
&\mu(x)=e^{-2cx}~,\\
&V_{1\gamma}(x)=c^2+\frac{4c}{\gamma e^{2cx}-\frac{1}{2c}}+\frac{2}{(\gamma e^{2cx}-\frac{1}{2c})^2}~,\\
&\Psi_{0\gamma}(x)=\frac{e^{-cx}}{\gamma -\frac{1}{2c} e^{-2cx}}~.
\end{align}
The plots of these functions for $c=1$ are displayed in Fig.~(\ref{Set6}). The critical $\gamma_s$ is
\begin{equation}\label{gsfree}
\gamma_s= -\frac{1}{2c}~.
\end{equation}
For strictly positive $c$, one should have $\gamma<-\frac{1}{2c}$ to avoid singularities, which again can be easily seen graphically.

If $c=0$, we have the trivial case given by $V_{2,1}(x)=0, \mu(x)=1, V_{1\gamma}(x)=0, \Psi_{0\gamma}(x)=0$.

\end{enumerate}

\section{Discussion and Possible Applications}\label{D}

The plots of this work allow us to shed light on a number of important features which went unnoticed in any of the previous works dealing with the one-parameter families of isospectral potentials. The first important aspect is that the shape of the integrating factor $\mu(x)$ is very important for the departure of the ground state solutions from the typical one corresponding to the particular Riccati solution. In addition, the plot of the integral of $\mu(x)$ up to an arbitrary taken $x$, which we denoted by $\gamma(x)$, is important for separating the regular and singular regions of parametric potentials and zero modes as well. Graphically, if we start from any point on the plot of $\gamma(x)$ and draw a horizontal line then the value of the intersection of that line with the ordinate axis will provide the value of the $\gamma$ parameter at which a singularity will occur. Thus, it is only when $\gamma(x)$ presents horizontal asymptotes that there is a possibility for a regular region of the parametric potentials and zero modes. For example, in the first case of asymmetric quadratic potentials, $\gamma(x)$ has an asymptote parallel to the positive half-line. That means that $\gamma(x)$ is practically constant for any $x\geq 1$. Horizontal asymptotes are important because one can take the values of $\gamma(\pm\infty)$ as $x$-independent quantities corresponding to $\gamma_s$ in equation (\ref{mh9d}). In the subcase of the harmonic oscillator, $\gamma(x)$ has two horizontal asymptotes along the two semi-axes and they are the borders for the singular region between them. For the plot of  $\gamma(x)$ in the Fresnel-like case, no asymptotes show up and therefore we get only singular potentials and zero modes. For example, one can see on the plot of $\gamma(x)$ for this case that $\gamma(5)\approx -2$. Thus, for $\gamma=-2$, the singularity occurs at $x\sim 5$ and for bigger negative $\gamma$'s the singularities are displaced further to bigger values of $x$.
Another interesting feature is that the last case of constant potential has an integrating factor very similar to the first case of asymmetric quadratic oscillator (and similar $\gamma(x)$ thereof) and one can furthermore think of the constant potential as a degenerate asymmetric parabolic case of such a type with the parabolic wings flatten down to horizontal asymptotes. On the other hand, one cannot take the constant potential as a degenerate case of the symmetric parabolic case from the viewpoint of the present parametric isospectral construction because the shape of the integrating factor is of the switching type in the latter case being closer to the Fresnel-like case if one ignores the periodic wiggles.

\begin{figure}[x!] 
\begin{center}
\resizebox*{0.3\textheight}{!}{
{\includegraphics{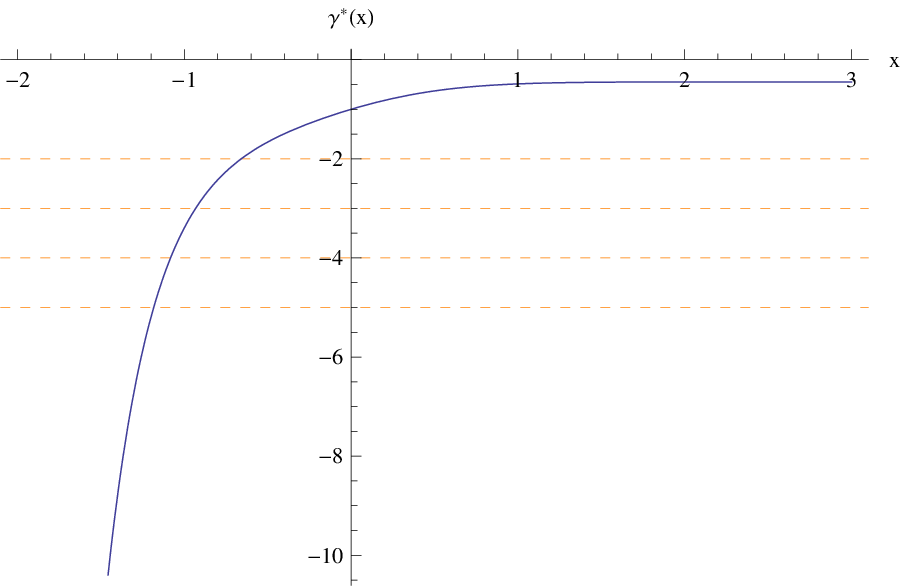}}} 
\resizebox*{0.3\textheight}{!}{
{\includegraphics{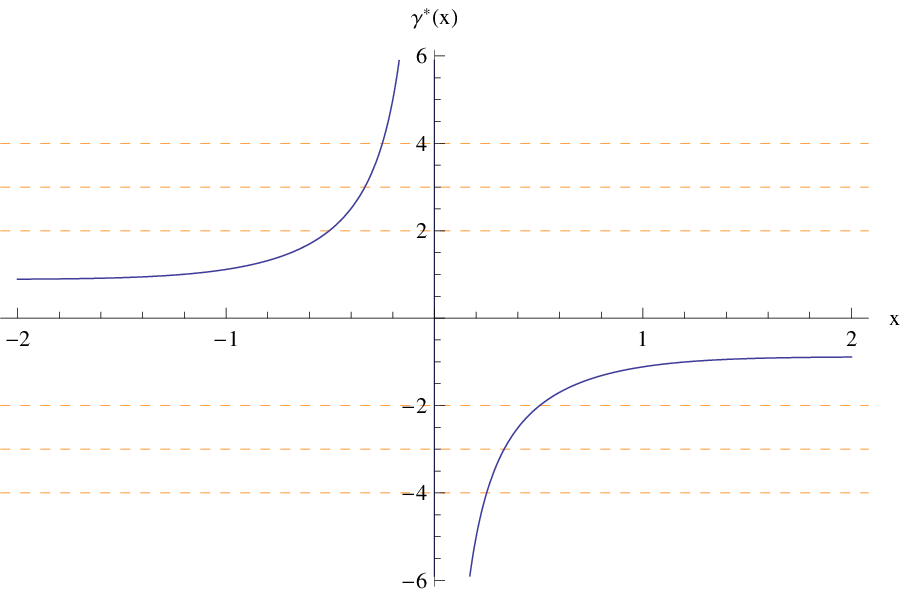}}} 
\resizebox*{0.3\textheight}{!}{
{\includegraphics{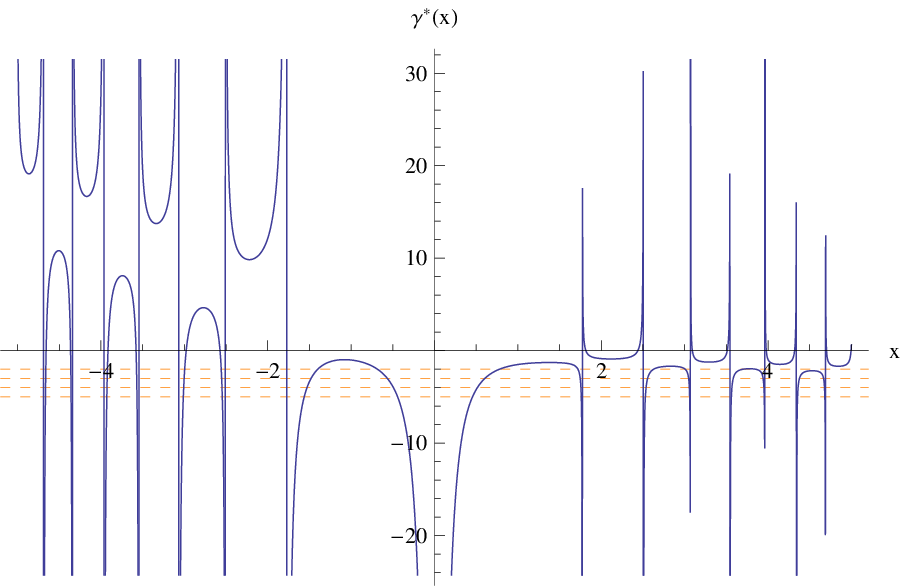}}} 
\resizebox*{0.35\textheight}{!}{
{\includegraphics{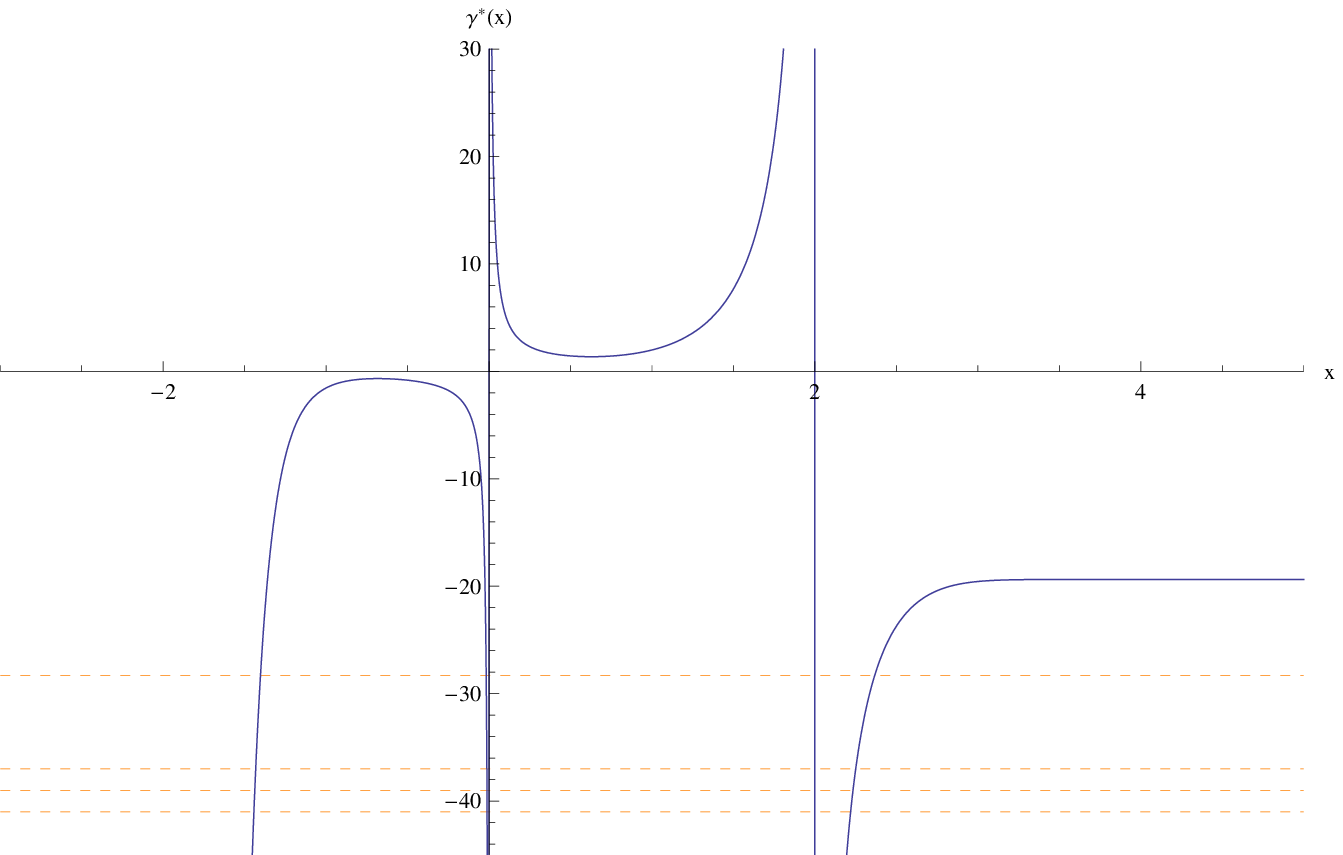}}} 
\resizebox*{0.3\textheight}{!}{
{\includegraphics{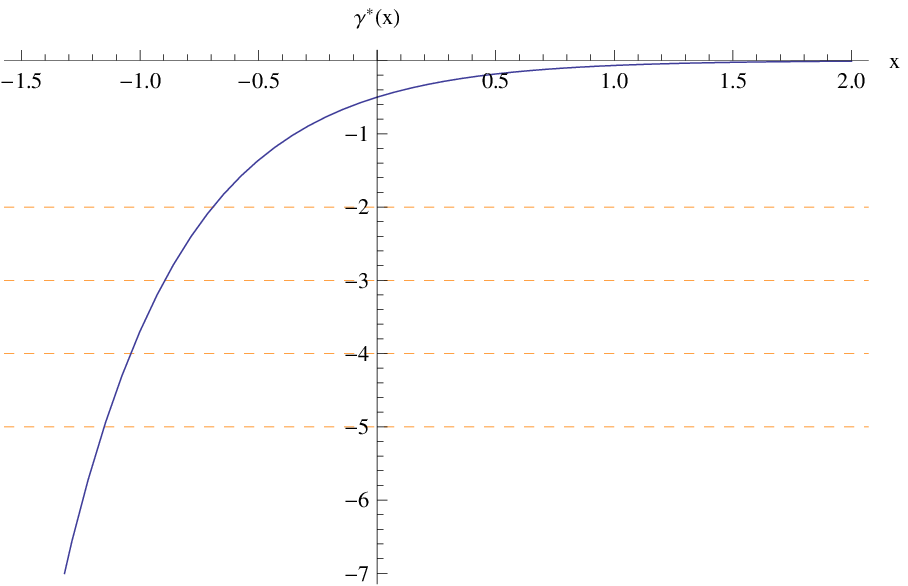}}} 
\caption{\textsl{(color online). Plots of $\gamma^*(x)$ for all the cases considered in the paper.
The intersections with horizontal lines of constant $\gamma$ define the positions of the maxima of the corresponding parametric zero modes $\overline{\Psi}_{0\gamma}^{2}$.
In the Fresnel-like case there are naturally more intersections.}}
\label{Set7}
\end{center}
\end{figure}

We also notice that the positions of the maxima of the normalized squared eigenfunctions $\overline{\Psi}_{0\gamma}^2$ can be found from the condition $\Phi_g=0$, i.e., from
\begin{equation}\label{14}
-F(x)=\frac{\mu(x)}{\gamma^*+\int_0^x \mu(x')dx'}~,
\end{equation}
and solving for $\gamma^*$ forces us to consider it as a function of $x$
\begin{equation}\label{15}
\gamma^*(x)=-\frac{\mu(x)+F(x)\int_0^x\mu(x')dx'}{F(x)}
\end{equation}
because the rhs of (\ref{15}) is a function of $x$.
Then the abscissas of the intersections of the horizontal lines $\gamma=const$ with $\gamma^*(x)$ provide the positions of the peaks of $\overline{\Psi}_{0\gamma}^2$.
This is shown in Fig.~(\ref{Set7}) for all the cases discussed in the paper.

\medskip

In the third case, which introduces a one-parameter quartic double well potential of asymmetric type,
we notice the very interesting fact that the parameter $\gamma$ acts as a control parameter for the height of the probability density in the two wells of the parametric potential. Indeed, one can have a higher height of the probability density of the particle in the shallower well for the range of $\gamma\in (\gamma_s,\gamma_c]$, where $\gamma_c$ is the value of $\gamma$ for which the heights of the peaks of $\overline{\Psi}_{0\gamma}^2$ in the two wells are equal. In our case, this happens for $\gamma\in (-19.3694,-28.33]$. This means that the particle localization can be manipulated through the parameter of the potential and can be even stronger in the smaller well than in the deeper one. This can lead to real applications once the physical and/or technological significance of the parameter $\gamma$ is clearly established. At the present time, there are hints that $\gamma$ may be related to the introduction of finite interval boundaries which lead to the modulation of the wavefunctions of the confined quantum system \cite{Monthus}.

\medskip

All the supersymmetric one-parameter isospectral potentials discussed in this paper have the same spectrum as the corresponding partner potentials $V_1$, which are given by $x^2+1- \frac{x}{\sqrt{x^2+1}}$, $x^2-1$, $\sin^2x^2-2x\cos x^2$, $[x(x-2)]^2-2x+2$, and $c^2$, respectively. Unfortunately, for the reader interested in the spectral problem, we have to say that this is not an easy task except for the harmonic oscillator and the constant potentials (the second and the last cases in the list).
The first potential is `harmonic-oscillator'-like with an additional non-harmonic contribution of the form $- \frac{x}{\sqrt{x^2+1}}$. Because of the latter term, the most direct way to find the spectrum is by numerical methods.
The Fresnel-like case looks as a sort of pseudo-periodic eigenvalue problems, which is hard to believe that can be analytical because even for pure periodic eigenvalue problems the Hill discriminant method is actually a numerical method \cite{kh2}. For the case of the quartic anharmonic potential only numerical methods are available. The last case of constant potential has a continuous spectrum and no discrete spectrum at all. However, from the continuous spectrum we used the zero mode solution corresponding to the integrating factor, which for the one-parameter isospectral potential turns into a normalized ground-state. If one uses other eigenstates from the continuous spectrum then what one can get are bound states in the continuum (also known as BICs) \cite{psp}.
Strictly speaking, the supersymmetric constructions based on the excited states lead only to singular potentials and wavefunctions because of the nodes of the wavefunctions. In recent years, however there have been found modified procedures based on excited states in which the singularities can be avoided \cite{mod1,mod2,mod3}.

\medskip

In conclusion, we presented here a detailed discussion of a function composition generalization of the well-known Mielnik construction of the one-parameter class of isospectral potentials based on zero mode solutions. Several interesting examples have been given to illustrate the procedure and the Mielnik parametric harmonic oscillator was included as a particular case. The applications could be similar to those already discussed in the literature for such types of potentials obtained from the non-composite functional form, i.e., to bound states in the continuum in quantum physics \cite{psp,pmi}, in photonic crystals \cite{pmr} and graded-index waveguides \cite{goyal}, as well as to generate soliton profiles \cite{kumar}. We also recall here the biological applications of the harmonic oscillator isospectral potential to the simulation of the H-bond in DNA, \cite{dfr}, and  the applications to Peyrard's microscopic model of nonlinear DNA dynamics \cite{agk} and travelling double-wells potentials in microtubules \cite{rosuetal}. Besides, very recently, the supersymmetric one-parameter isospectral potentials have been used with very interesting goals in two important areas:

(i) In the context of optical solitons of the one-dimensional nonlinear Schr\"odinger equation, Yang \cite{yang} used one-parameter isospectral potentials with complex parameter $\gamma$ to study under what conditions soliton families parametrized by their propagation constants can bifurcate out from the linear guided modes which correspond to the discrete real eigenvalues of the corresponding linear Schr\"odinger spectral problem. In particular, he used the parametric harmonic oscillator zero mode (\ref{210-4}) to show that only when $\gamma$ is real or purely imaginary, which means parity-time ($PT$) symmetric potentials, the bifurcation of solitons occurs.

(ii) Curtright and Zachos \cite{CZ} used the one-parameter isospectral construction in the momentum space for a case involving multi-valued (branched) Hamiltonians based on the Riccati particular solution $F(p)=\sqrt{p}$, which in the position representation may be considered as a subcase of our first example. Any isospectral Hamiltonian from the one-parameter family can be in this case the other branched Hamiltonian which although exists in the same Hilbert space as the non-parametric Hamiltonian connects with it only at $p=\infty$. This looks a very exotic application, but according to Shapere and Wilczek \cite{sw1,w2,sw3}  `many-worlds' systems with branched Hamiltonians are encountered in a broad class of theoretical models and an experimental setup with an ion ring in a cylindrical trap has been already proposed \cite{tli}. We have here another possible application of the function composition formulation of the Riccati solutions in the sense that it can help in classifying the various types of supersymmetric branched Hamiltonians.

\bigskip
\medskip

%
{\bf Appendix. The normalization constant of the parametric zero modes}\\

Except for the Fresnel oscillatory case, all the other parametric families of potentials presented in this work support bound (localized) zero modes for which not only the regularity of the potentials is important but also to endow them with the normalization constant. For completeness, we present this issue in this Appendix.

We can always write a one-parameter zero mode, say $u$, in the form
\begin{equation}\label{a1}
 u(x)=\frac{u_0(x)}{\gamma+\int_l^xu_0^2(x')dx'}~,
 \end{equation}
where $l$, the lower limit of the integral, is naught for half-line problems and $-\infty$ for full-line problems, while $u_0$, the original zero mode of the nonparametric potential, is assumed normalized, i.e., $\int_{l}^{\infty}u_0^2(x)dx=1$. 
To get $u$ normalized, we need to write the normalization condition
\begin{equation}\label{a2}
\int_{l}^{\infty}N^2\frac{u_0^2dx}{(\gamma+\int_l^xu_0^2(x')dx')^2}=1
\end{equation}
and introduce further the change of variable $X(x)=\int_l^x u_0^2(x')dx'$ which turns (\ref{a2}) into the form
\begin{equation}\label{a3}
\int_0^1N^2\frac{dx}{(\gamma+x)^2}=1~.
\end{equation}
The final result is $N=\sqrt{\gamma(\gamma+1)}$. Thus, $\gamma$ should not be in the interval $[-1,0]$ in quantum mechanical applications.


\renewcommand{\baselinestretch}{1.0}

\renewcommand{\baselinestretch}{1.0}

\renewcommand{\baselinestretch}{1.0}

\renewcommand{\baselinestretch}{1.0}
\begin{figure}[x!] 
\begin{center}
\resizebox*{0.3\textheight}{!}{
{\includegraphics{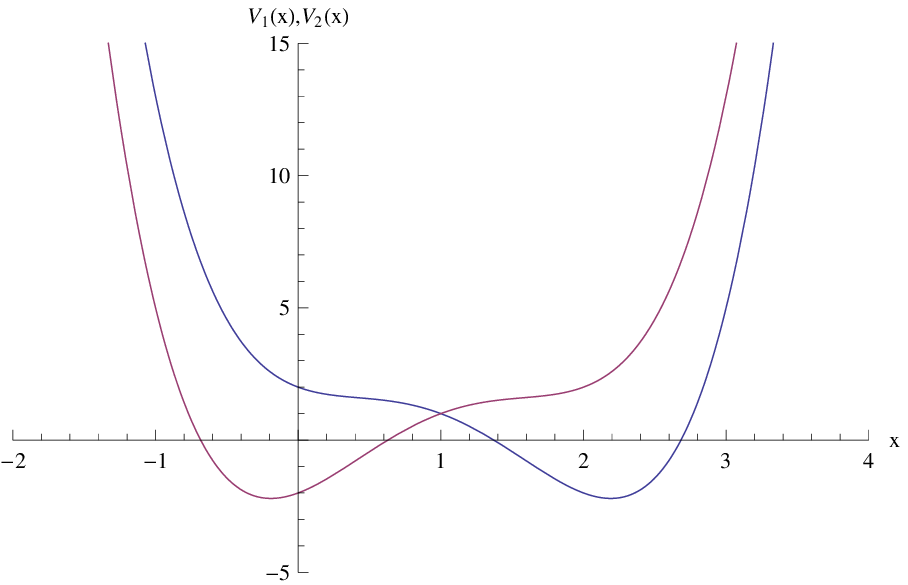}}}
\resizebox*{0.3\textheight}{!}{
{\includegraphics{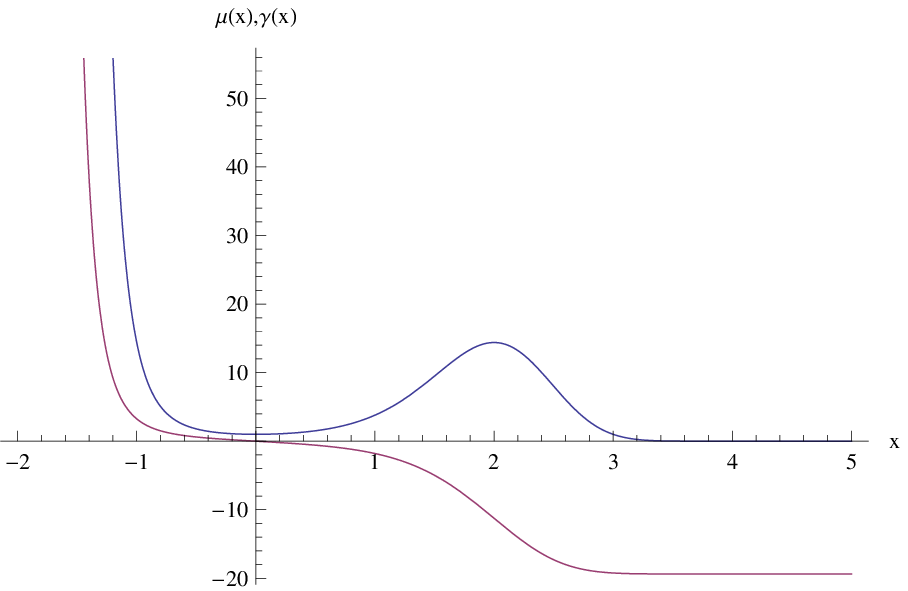}}}
\resizebox*{0.3\textheight}{!}{
{\includegraphics{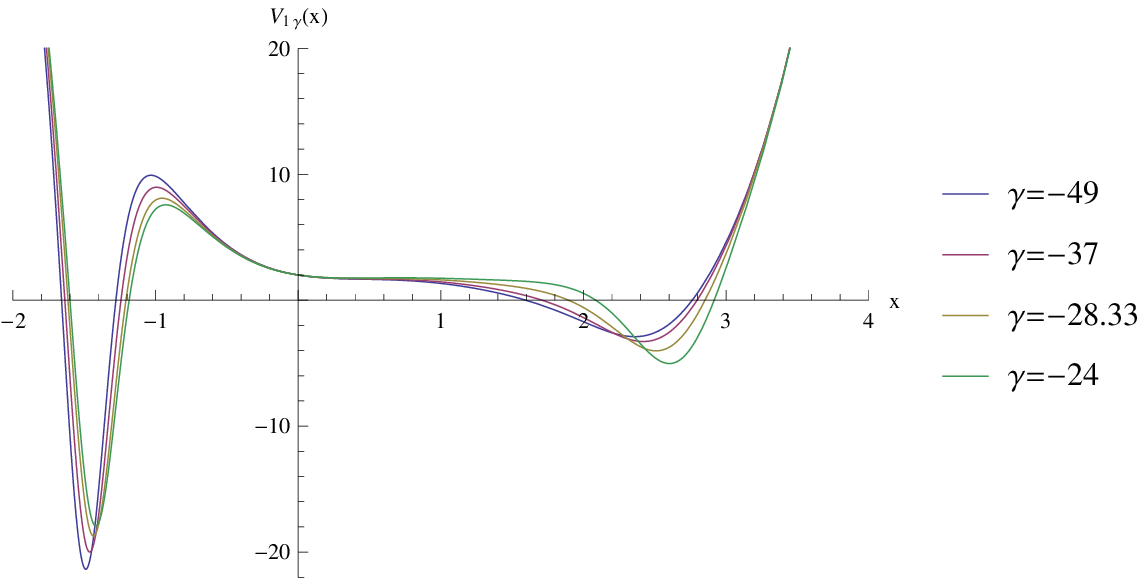}}}
\resizebox*{0.3\textheight}{!}{
{\includegraphics{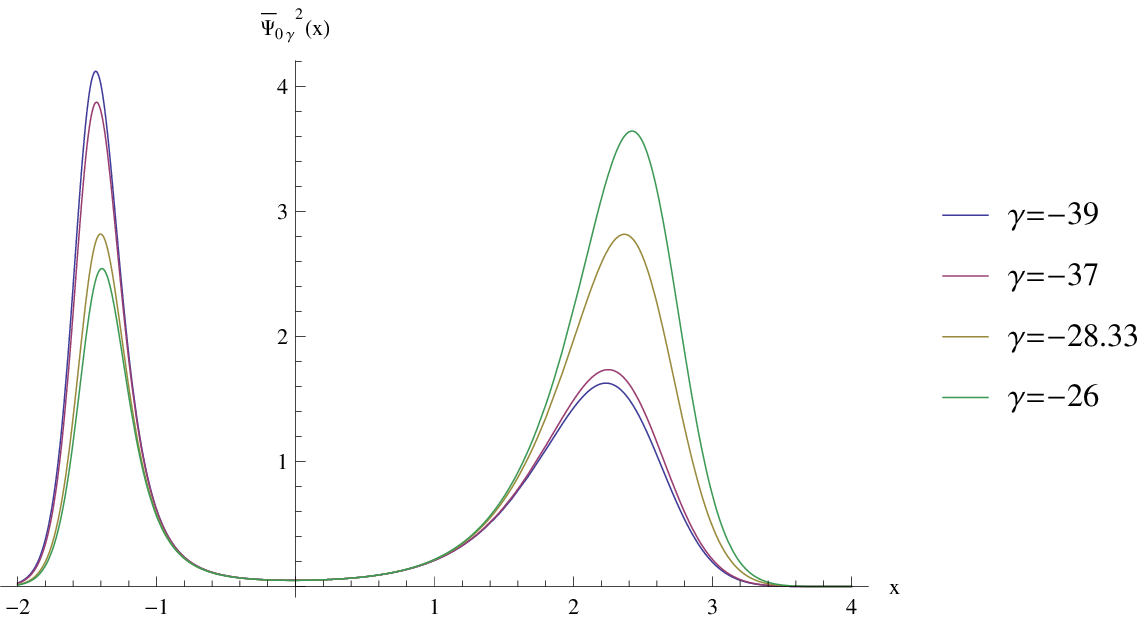}}}
\resizebox*{0.3\textheight}{!}{
{\includegraphics{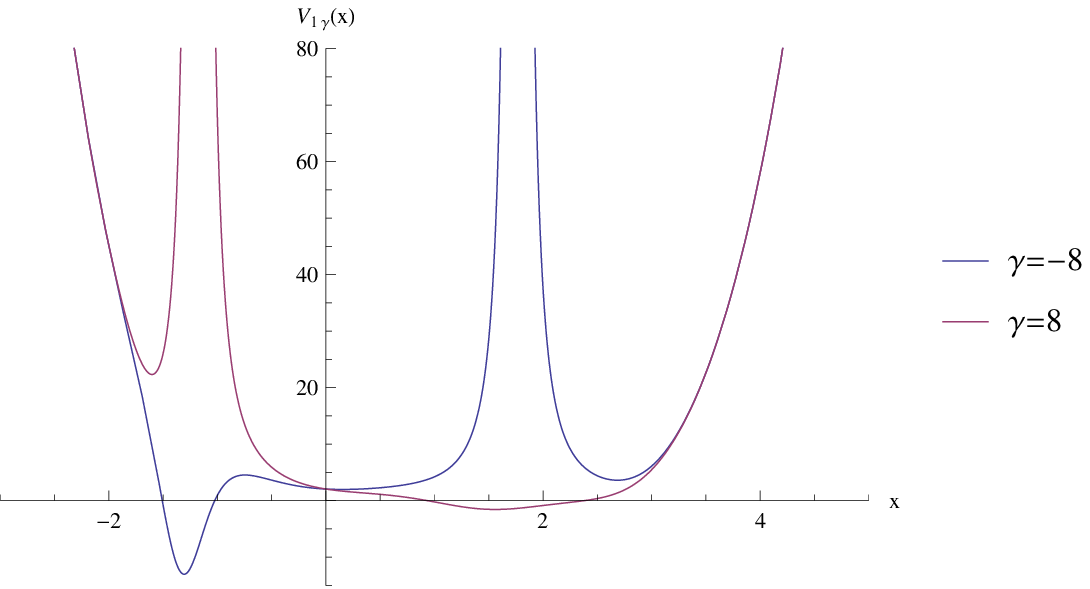}}}
\resizebox*{0.3\textheight}{!}{
{\includegraphics{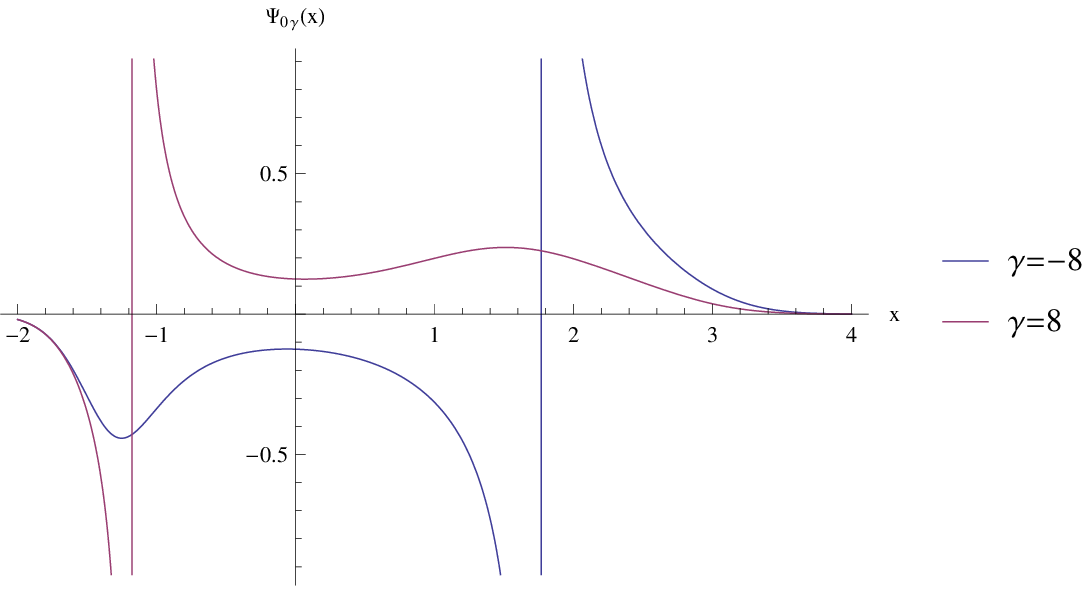}}}
\caption{\textsl{(color online). The partner asymmetric quartic potentials, $V_1$ (blue) and $V_2$ (red);  integrating factor $e^{-\frac{2}{3}x(x^2-3x)}$ (blue) and $\gamma(x;-3)$ (red);   one-parameter potentials, $V_{1\gamma}$, and normalized ground state squared eigenfunctions, $\overline{\Psi}_{0\gamma}^{2}$, for $\gamma=-49,-37,-28.33 $ and $-26$;  unbounded potentials and eigenfunctions for $\gamma =\pm 8$.}}
\label{SetQ}
\end{center}
\end{figure}

\renewcommand{\baselinestretch}{1.0}
\begin{figure}[x!] 
\begin{center}
\resizebox*{0.3\textheight}{!}{
{\includegraphics{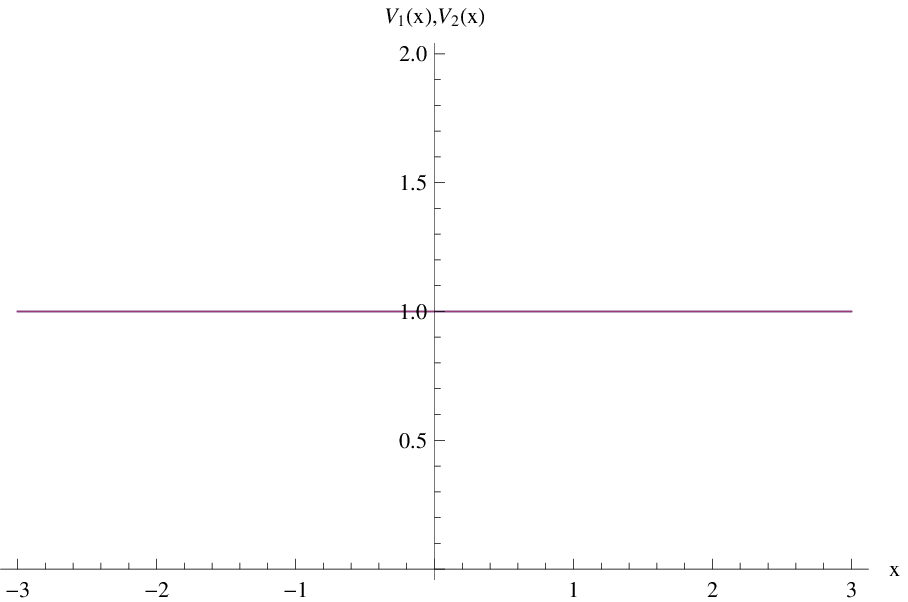}}}
\resizebox*{0.3\textheight}{!}{
{\includegraphics{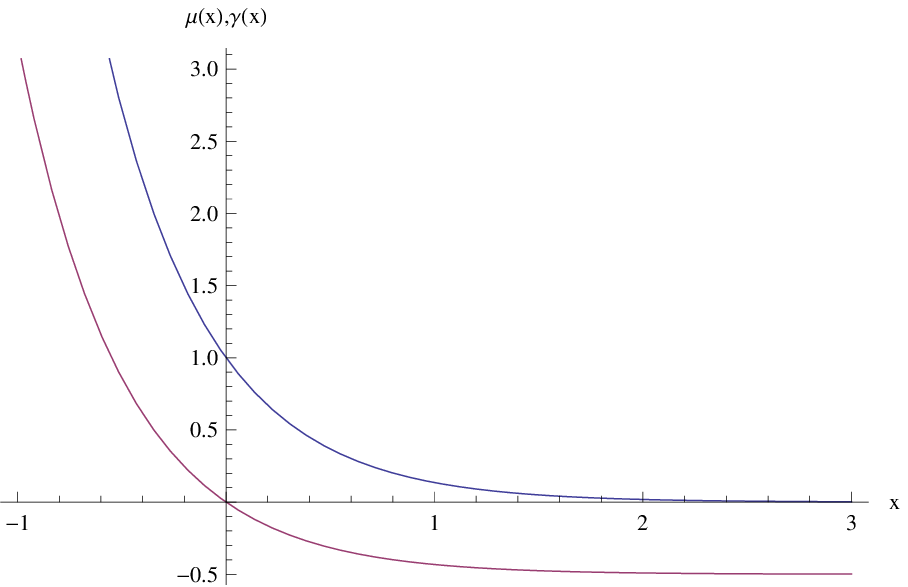}}}
\resizebox*{0.3\textheight}{!}{
{\includegraphics{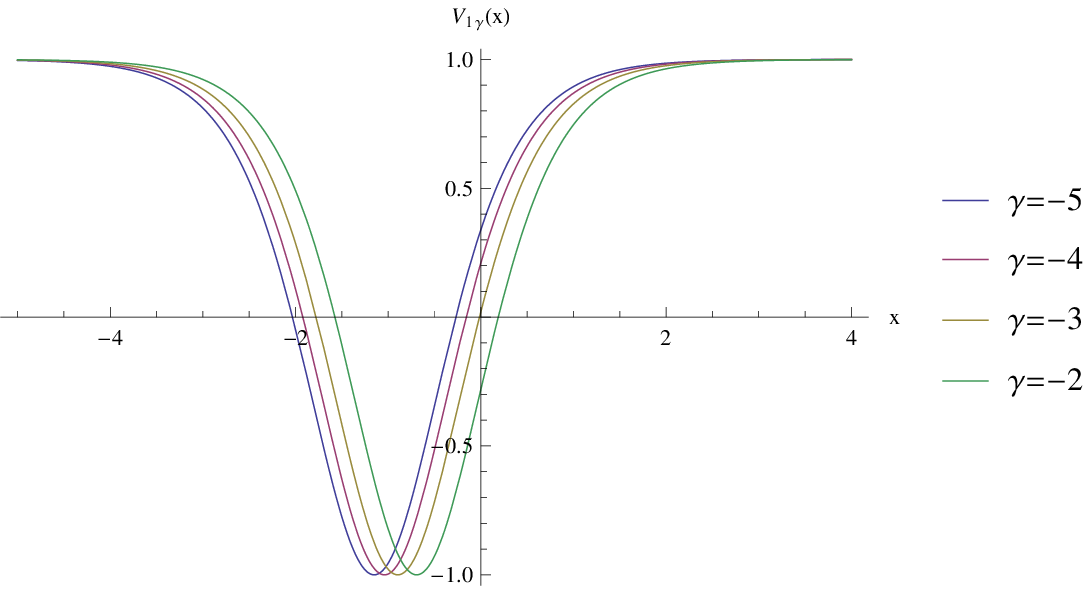}}}
\resizebox*{0.3\textheight}{!}{
{\includegraphics{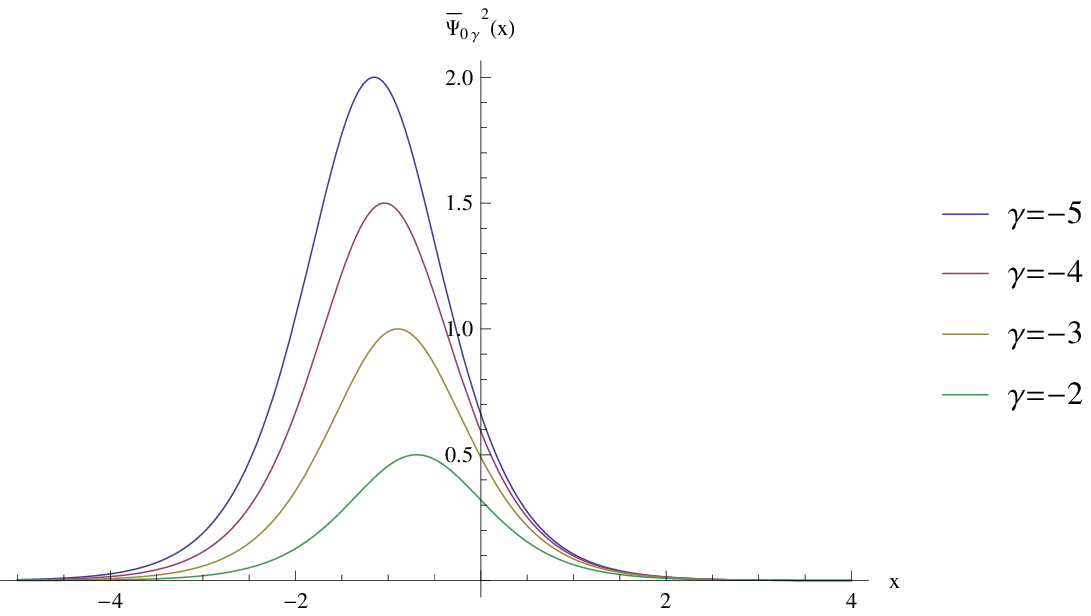}}}
\resizebox*{0.3\textheight}{!}{
{\includegraphics{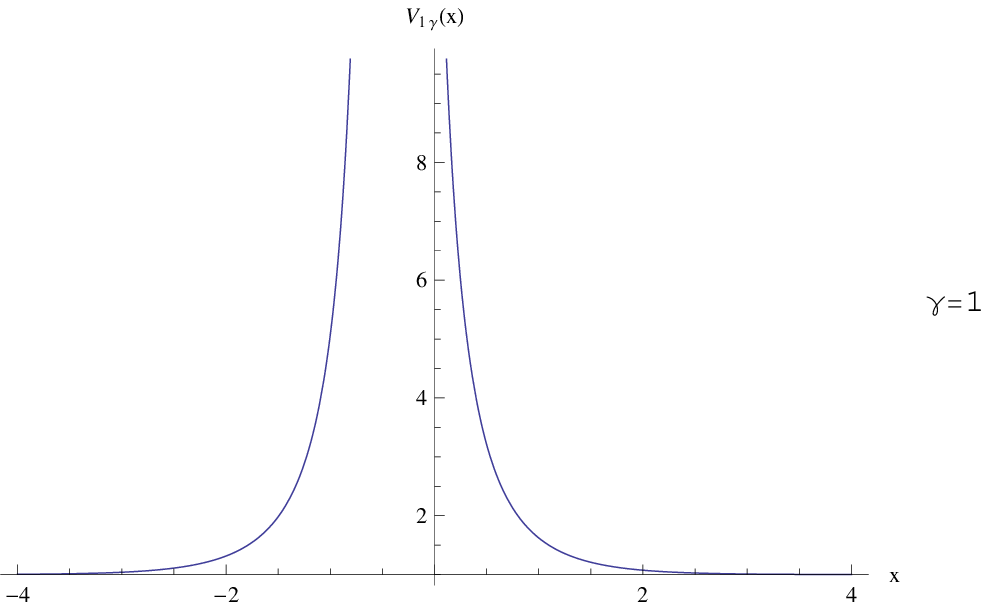}}}
\resizebox*{0.3\textheight}{!}{
{\includegraphics{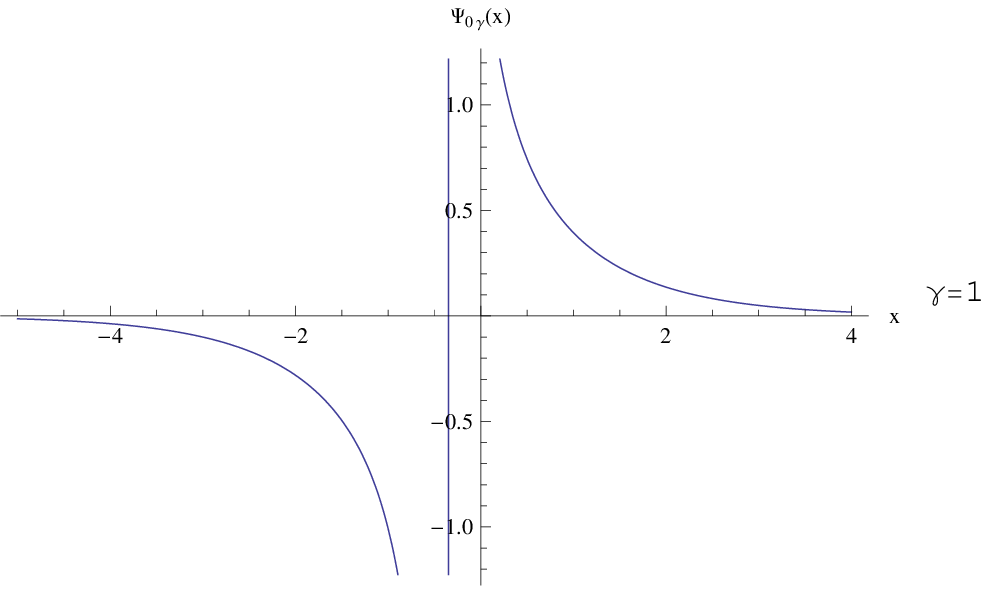}}}
\caption{\textsl{(color online). The constant potential case for $c=1$: The constant partner potentials, $V_1$ and $V_2$;  integrating factor $e^{-2x}$ (blue) and $\gamma(x)$ (red);  one-parameter potentials $V_{1\gamma}$ and normalized squared zero modes $\overline{\Psi}_{0\gamma}^{2}$ for $\gamma=-5,-4,-3$, and $-2$; singular potential and zero mode for $\gamma=1$.}}
\label{Set6}
\end{center}
\end{figure}
%
\renewcommand{\baselinestretch}{1.0}


{\small

\begin{thebibliography}{999}

\bibitem{FM0} L. Infeld, T.D. Hull,
{\textit The factorization method},
Rev. Mod. Phys. 23, 21-68 (1951).

\bibitem{FM1} B. Mielnik, O. Rosas-Ortiz,
{\textit Factorization: little or great algorithm ?},
J. Phys. A: Math. Gen. 37, 10007-10035 (2004).

\bibitem{FM2} S.-H. Dong, {\textit Factorization Method in Quantum Mechanics}, Springer, Dordrecht, The Netherlands, 2007.

\bibitem{witt} E. Witten,
 {\textit Dynamical breaking of supersymmetry},
Nucl. Phys. B 185, 513-554 (1981).

\bibitem{hr86} R. Haymaker, A.R.P. Rau,
 {\textit Supersymmetry in quantum mechanics},
Am. J. Phys. 54, 928-936 (1986).

\bibitem{bijker} R. Bijker,
 {\textit Supersymmetry in nuclear physics},
J. Phys.: Conf. Series 237, 012005 (2010).

\bibitem{M} B. Mielnik,
 {\textit Factorization method and new potentials with the oscillator spectrum},
J. Math. Phys. 25, 3387-3389 (1984).

\bibitem{F} D.J. Fern\'andez C.,
 {\textit New Hydrogen-like potentials},
Lett. Math. Phys. 8, 337-343 (1984).

\bibitem{N} M.M. Nieto,
 {\textit Relationship between supersymmetry and the inverse method in quantum mechanics},
Phys. Lett. B 145, 208-2010 (1984).

\bibitem{B1} B. Bagchi,  {\textit Supersymmetry in Quantum and Classical Mechanics}, sections 5.6 and 5.7, Chapman and Hall/CRC, Boca Raton, Fl., 2001.

\bibitem{B2} F. Cooper, A. Khare, U. Sukhatme,  {\textit Supersymmetry in Quantum Mechanics}, sections 6.1 and 6.2, World Scientific, Singapore, 2001.

\bibitem{TR} D. Baye, J.-M. Sparenberg, A.M. Pupasov-Maksimov, B.F. Samsonov,  {\textit Single- and coupled-channel inverse scattering with supersymmetric transformations}, arXiv:1401.0439.

\bibitem{MH} M.K. Mak, T. Harko,
 {\textit New further integrability cases for the Riccati equation},
Appl. Math. Comput. 219, 7465-7471 (2013). arXiv:1301.5720.


\bibitem{r98} H.C. Rosu,  {\textit Short survey of Darboux transformations}, in: Symmetries in Quantum Mechanics and Quantum Optics, A. Ballesteros et al. (Eds), Serv. de Publ. Univ. Burgos, Burgos 1999, pp. 301-315. arXiv:quant-ph/9809056.

\bibitem{Monthus} C. Monthus, G. Oshanin, A. Comtet, S.F. Burlatsky,
 {\textit Sample-size dependence of the ground-state energy in a 1D localization problem},
Phys. Rev. E 54, 231-242 (1996).

\bibitem{kh2} K.V. Khmelnytskaya, H.C. Rosu,
 {\textit Spectral parameter power series representation for Hill's discriminant},
Ann. Phys. 325, 2512-2521 (2010).

\bibitem{psp} J. Pappademos, U. Sukhatme, A. Pagnamenta,
 {\textit Bound states in the continuum from supersymmetric quantum mechanics},
Phys. Rev. A 48, 3525-3531 (1993).

\bibitem{mod1} M.S. Berger, N.S. Ussembayev,
 {\textit Isospectral potentials from modified factorization},
Phys. Rev. A 82, 022121 (2010).

\bibitem{mod2} D. Dutta, P. Roy,
 {\textit Generalized factorization and isospectral potentials},
Phys. Rev. A 83, 054102 (2011).

\bibitem{mod3} B. Midya,
 {\textit Nonsingular potentials from excited state factorization of a quantum system with position-dependent mass},
J. Phys. A: Math. Theor. 44, 435306 (2012).

\bibitem{pmi} J.S. Petrovi\'c, V. Milanovi\'c, Z. Ikoni\'c,
 {\textit Bound states in continuum of complex potentials generated by supersymmetric quantum mechanics},
Phys. Lett. A 300, 595-602 (2002).

\bibitem{pmr} N. Prodanovi\'c, V. Milanovi\'c, J. Radovanovi\'c,
 {\textit Photonic crystals with bound states in the continuum and their realization by an advanced digital grading method},
J. Phys. A: Math. Theor. 42, 415304 (2009).

\bibitem{goyal} A. Goyal, R. Gupta, S. Loomba, C.N. Kumar,
 {\textit Riccati parametrized self-similar waves in tapered graded-index waveguides},
Phys. Lett. A 376, 3454-3457 (2012).

\bibitem{kumar} C.N. Kumar,
 {\textit Isospectral Hamiltonians: generation of the soliton profile},
J. Phys. A: Math. Gen. 20, 5397-5401 (1987).

\bibitem{dfr} E. Drigo-Filho, J.R. Ruggiero,
 {\textit H-bond simulation in DNA using a harmonic oscillator isospectral potential},
Phys. Rev. E 56, 4486-4488 (1997).

\bibitem{agk} W. Alka, A. Goyal, C.N. Kumar,
 {\textit Nonlinear dynamics of DNA - Riccati generalized solitary wave solutions},
Phys. Lett. A 375, 480-483 (2011).

\bibitem{rosuetal} H.C. Rosu, J.M. Mor\'an-Mirabal, O. Cornejo,
 {\textit One-parameter nonrelativistic supersymmetry for microtubules},
Phys. Lett. A 310, 353-356 (2003).

\bibitem{yang} J. Yang,
 {\textit Necessity of PT symmetry for soliton families in 1D complex potentials},
Phys. Lett. A 378, 367-373 (2014). arXiv: 1310.4490.

\bibitem{CZ}  T.L. Curtright, C.K. Zachos,
 {\textit Branched Hamiltonians and supersymmetry},
arXiv: 1311.6147v2.

\bibitem{sw1} A. Shapere, F. Wilczek,
 {\textit Classical time crystals},
Phys. Rev. Lett. 109, 160402 (2012).

\bibitem{w2} F. Wilczek,
 {\textit Quantum time crystals},
Phys. Rev. Lett. 109, 160401 (2012).

\bibitem{sw3}  A. Shapere, F. Wilczek,
 {\textit Branched quantization},
Phys. Rev. Lett. 109, 200402 (2012).

\bibitem{tli} T. Li, Z-X. Gong, Z-Q. Yin, H. T. Quan, X. Yin, P. Zhang, L-M. Duan, X. Zhang,
 {\textit Space-time crystals of trapped ions},
Phys. Rev. Lett. 109, 163001 (2012).

\end{thebibliography}

}
\end{document}